\documentclass[sigconf, screen]{acmart}

\AtBeginDocument{%
  \providecommand\BibTeX{{%
    \normalfont B\kern-0.5em{\scshape i\kern-0.25em b}\kern-0.8em\TeX}}}

\setcopyright{acmcopyright}
\acmPrice{15.00}
\acmDOI{10.1145/3445814.3446747}
\acmYear{2021}
\copyrightyear{2021}
\acmSubmissionID{asplos21main-p898-p}
\acmISBN{978-1-4503-8317-2/21/04}
\acmConference[ASPLOS '21]{Proceedings of the 26th ACM International Conference on Architectural Support for Programming Languages and Operating Systems}{April 19--23, 2021}{Virtual, USA}
\acmBooktitle{Proceedings of the 26th ACM International Conference on Architectural Support for Programming Languages and Operating Systems (ASPLOS '21), April 19--23, 2021, Virtual, USA}

\usepackage[normalem]{ulem}

\usepackage{amsmath}
\usepackage{tabularx}
\usepackage{lscape}
\usepackage{mathtools}
\usepackage{booktabs}
\usepackage{color}
\usepackage{xcolor}
\usepackage{algpseudocode}
\usepackage{algorithm}

\DeclareMathOperator*{\argmin}{arg\,min}
\usepackage{enumitem}

\usepackage[toc,page]{appendix}
\usepackage{xcolor}
\usepackage{graphicx}

\usepackage{balance}
\begin{document}

\title{Defensive Approximation: Securing CNNs using Approximate Computing}
\author{Amira Guesmi}
\email{amira.guesmi@stud.enis.tn}
\orcid{0002-8992-7958}
\affiliation{%
  \institution{ENIS, University of Sfax}
  \country{Tunisia}
}

\author{Ihsen Alouani}
\email{Ihsen.Alouani@uphf.fr}
\affiliation{%
  \institution{IEMN, Polytechnic University Hauts-De-France}
  \country{France}}

\author{Khaled N. Khasawneh}
\email{kkhasawn@gmu.edu}
\affiliation{%
  \institution{George Mason University}
  \country{USA}
}

\author{Mouna Baklouti}
\affiliation{%
 \institution{ENIS, University of Sfax}
 \country{Tunisia}}

\author{Tarek Frikha}
\affiliation{%
 \institution{ENIS, University of Sfax}
 \country{Tunisia}}

\author{Mohamed Abid}
\affiliation{%
 \institution{ENIS, University of Sfax}
 \country{Tunisia}}

\author{Nael Abu-Ghazaleh}
\affiliation{%
  \institution{University of California Riverside}
  \country{USA}}

\renewcommand{\shortauthors}{Amira Guesmi, Ihsen Alouani, Khaled N. Khasawneh, Mouna Baklouti, Tarek Frikha, Mohamed Abid, and Nael Abu-Ghazaleh}

\begin{abstract}
In the past few years, an increasing number of machine-learning and deep learning structures, such as Convolutional Neural Networks (CNNs), have been applied to solving a wide range of real-life problems.  However, these architectures are vulnerable to adversarial attacks: inputs crafted carefully to force the system output to a wrong label. Since machine-learning is being deployed in safety-critical and security-sensitive domains, such attacks may have catastrophic security and safety consequences.  
In this paper, we propose for the first time to use hardware-supported approximate computing to improve the robustness of machine learning classifiers. 
We show that our approximate computing implementation achieves robustness across a wide range of attack scenarios. Specifically, we show that successful adversarial attacks against the exact classifier have poor transferability to the approximate implementation. The transferability is even poorer for the black-box attack scenarios, where adversarial attacks are generated using a proxy model. Surprisingly, the robustness advantages also apply to white-box attacks where the attacker has unrestricted access to the approximate classifier implementation: in this case, we show that substantially higher levels of adversarial noise are needed to produce adversarial examples. 
Furthermore, our approximate computing model maintains the same level in terms of classification accuracy, does not require retraining, and reduces resource utilization and energy consumption of the CNN.
We conducted extensive experiments on a set of strong adversarial attacks; We empirically show that the proposed implementation increases the robustness of a LeNet-5 and an Alexnet CNNs by up to $99\%$ and $87\%$, respectively for strong transferability-based attacks along with up to $50\%$ saving in energy consumption due to the simpler nature of the approximate logic. We also show that a white-box attack requires a remarkably higher noise budget to fool the approximate classifier, causing an average of $4
~dB$ degradation of the PSNR of the input image relative to the images that succeed in fooling the exact classifier.
\end{abstract}




\maketitle

\section{Introduction}

Convolutional neural networks (CNNs) and other deep learning structures  provide state-of-the-art performance in many application domains, such as computer vision \cite{simonyan2014deep,redmon2016yolo9000}, natural language processing \cite{deng2018deep}, robotics \cite{pierson2017deep}, autonomous driving \cite{al2017deep}, and healthcare \cite{miotto2018deep}. With the rapid progress in CNN's development and deployment, they are increasing concerns about their vulnerability to adversarial attacks: maliciously designed imperceptible perturbations injected within the data that cause CNNs to misclassify.   Adversarial attacks have been demonstrated in real-world scenarios \cite {phy9,phy10,neuroattack}, making this vulnerability a serious threat to safety-critical and other applications that rely on CNNs. 


Several software-based defenses have been proposed against Adversarial Machine Learning (AML) attacks~\cite{distillation_SP, nayebi2017biologically, na2017cascade}, but more advanced attack strategies~\cite{CW, DLres} also continue to evolve that demonstrate vulnerability of some of these defenses. Moreover, many of the proposed defenses introduce substantial overheads to either the training or the inference operation of CNNs~\cite{xie2019feature, na2017cascade}.  These overheads increase the computational requirements of these systems, which is already a significant challenge driving substantial research into algorithmic and hardware techniques to improve CNN performance and energy-efficiency~\cite{reviewDL}.    
Thus, finding new approaches to harden systems against AML without heavy overheads in both design-time and run-time is an area of substantial need.

In this paper, we propose a new hardware based approach to improve the robustness of machine learning (ML) classifiers.  Specifically, we show that Approximate Computing (AC),  designed to improve the performance and power consumption of algorithms and processing elements,  can substantially improve CNN robustness to AML.  
Our technique, which we call \emph{defensive approximation} (DA), substantially enhances the robustness of CNNs to adversarial attacks. 
We show that for a variety of attack scenarios, and utilizing a range of algorithms for generating adversarial attacks, DA provides substantial robustness even under the assumptions of a powerful attacker with full access to the internal classifier structure.   Importantly, DA does not require retraining or fine-tuning, allowing pre-trained models to benefit from its robustness and performance advantages by simply replacing the exact multiplier implementations with approximate ones.   The approximate classifier achieves similar accuracy to the exact classifier for Lenet-5 and Alexnet. 

DA also benefits from the conventional advantages of AC, resulting in a less complex design that is both faster and more energy efficient.   Other defenses such as Defensive Quantization (DQ)~\cite{DQ} also provide energy efficiency benefits in addition to robustness.  However, we show that because it is input-dependent DA achieves twice higher robustness to attacks than DQ.

We carry out several experiments to better understand the robustness advantages of DA.  We show that the unpredictable and input-dependent variations introduced by AC improve the CNN resilience to adversarial perturbations. Experimental results show that DA has a confidence enhancement impact on non-adversarial examples; we believe that this is due to our AC multiplier which adds input dependent approximation with generally higher magnitude at large multiplication values. In fact, the AC-induced noise in the convolution layer is shown to be higher in absolute value when the input matrix is highly correlated to the convolution filter, and by consequence highlights further the features. This observation at the feature map propagates through the model and results in enhanced classification confidence, i.e., the difference between the $1^{st}$ class and the "runner-up". Intuitively and as shown by prior work~\cite{smooth}, enhancing the confidence furthers the classifier's robustness.
At the same time, we observe negligible accuracy loss compared to a conventional CNN implementation on non-adversarial inputs while providing considerable power savings.

In summary, the contributions of the paper are:
\begin{itemize}
    \item We build an aggressively approximate floating point multiplier that injects data-dependent noise within the convolution calculation. Subsequently, we used this approximate multiplier to implement an approximate CNN hardware accelerator (Section \ref{sec:AppxCNN}).  
    \item To the best of our knowledge, we are the first to leverage AC to enhance CNN robustness to adversarial attacks without the need for re-training, fine-tuning, nor input pre-processing. We investigate the capacity of AC to help defending against adversarial attacks in Section \ref{sec:secu}.
    \item We empirically show that the proposed approximate implementation reduces the success rate of adversarial attacks by an average of 87\% and 71.5\% in Lenet-5 and Alexnet CNNs respectively.
    \item We evaluate the approximate classifiers against powerful attackers with white-box access.  We observe that attackers require substantially higher adversarial perturbations to fool the approximate classifier.
    

\end{itemize}

\noindent 
We believe that DA is highly practical; it can be deployed without retraining or fine-tuning, achieving comparable classification performance to exact classifiers.  In addition to security advantages, DA {\em improves} performance by reducing latency and energy making it an attractive choice even in Edge device settings (Appendix \ref{sec:perf_impl}).

\section{Background}
This section first presents an overview of adversarial attacks followed by introducing AC.  



\subsection{Adversarial Attacks}

Deep learning techniques gained popularity in recent years and are now deployed even in safety-critical tasks, such as recognizing road signs for autonomous vehicles \cite{signs}. Despite their effectiveness and popularity, CNN-powered applications are facing a critical challenge – adversarial attacks. Many studies \cite{vulnerable, fgsm, He, neuroattack} have shown that CNNs are vulnerable to carefully crafted inputs designed to fool them, very small imperceptible perturbations added to the data can completely change the output of the model. In computer vision domain, these adversarial examples are intentionally generated images that look almost exactly the same as the original images, but can mislead the classifier to provide wrong prediction outputs. Other work \cite{noneed} claimed that adversarial examples are not a practical threat to ML in real-life scenarios. However, physical adversarial attacks have recently been shown to be effective against CNN based applications in real-world \cite{phys}.

\noindent 
{\bf Minimizing Injected Noise:} Its essential for the adversary to minimize the added noise to avoid detection.  
For illustration purposes, consider a CNN used for image classification.  More formally, given an original input image $x$ and a target classification model $ f() ~s.t. ~ f(x) = l $, the problem of generating an adversarial example $x^*$ can be formulated as a constrained optimization \cite{pbform}:

\begin{equation}
\label{eq:adv}
     \begin{array}{rlclcl}
        x^* = \displaystyle \argmin_{x^*}  \mathcal{D}(x,x^*),  
         s.t.  ~ f(x^*) = l^*,  ~ l \neq l^*
\end{array}
\end{equation}

Where $\mathcal{D}$ is the distance metric used to quantify the similarity between two images and the goal of the optimization is to minimize this added noise, typically to avoid detection of the adversarial perturbations. $l$ and $l^*$ are the two labels of $x$ and $x^*$, respectively:  $x^*$ is considered as an adversarial example if and only if the label of the two images are different ($ f(x) \neq  f(x^*) $) and the added noise is bounded ($\mathcal{D}(x,x^*) < \epsilon $ where $\epsilon \geqslant 0 $).

\noindent
{\bf Distance Metrics:}
The adversarial examples and the added perturbations should be visually imperceptible by humans. Since it is hard to model humans' perception, researchers proposed three metrics to approximate humans' perception of visual difference, namely $L_0$, $L_2$, and $L_ \infty$ \cite{CW}. These metrics are special cases of the $L_p$ norm: 
\begin{equation}
    \left\|x\right\|_p = \left( \sum^{n}_{i = 1} \left |x_i \right | ^{p} \right)^{\frac{1}{p}}
\end{equation}
These three metrics focus on different aspects of visual significance. $L_0$ counts the number of pixels with different values at corresponding positions in the two images. $L_2$ measures the Euclidean distance between the two images $x$ and $x^*$. $L_ \infty$ measures the maximum difference for all pixels at corresponding positions in the two images.

The consequences of an adversarial attack can be dramatic.  For example, misclassification of a stop sign as a yield sign or a speed limit sign could lead to material and human damages. Another possible situation is when using CNNs in financial transactions and automatic bank check processing -- using handwritten character recognition algorithms to read digits from bank cheques or using neural networks for amount and signature recognition \cite{cheque}. An attacker could easily fool the model to predict wrong bank account numbers or amount of money or even fake a signature. A dangerous situation, especially with such large sums of money at stake.


\subsection{Approximate Computing}

The speed of new generations of computing systems, from embedded and mobile devices to servers and computing data centers, has been drastically climbing in the past decades. This development was made possible by the advances in integrated circuits (ICs) design and driven by the increasingly high demand for performance in the majority of modern applications. However, this development is physically reaching the end of Moore’s law, since TSMC and Samsung are releasing 5 $nm$ technology \cite{tsmc_moore}. On the other hand, a wide range of modern applications is inherently fault-tolerant and may not require the highest accuracy. This observation has motivated the development of approximate computing (AC), a computing paradigm that trades power consumption with accuracy. The idea is to implement inexact/AC elements that consume less energy, as far as the overall application tolerates the imprecision level in computation.
This paradigm has been shown promising for inherently fault-tolerant applications such as deep/ML, big data analytics, and signal processing. Several AC techniques have been proposed in the literature and can be classified into three main categories based on the computing stack layer they target: software, architecture, and circuit level~\cite{Survey}. 

In this paper, we consider AC for a \emph{totally new objective}; enhancing CNNs robustness to adversarial attacks, without losing the initial advantages of AC.

\section{Threat Model}
We assume an attacker attempting to conduct adversarial attacks to fool a classifier in a variety of attack scenarios.

\subsection{Adversary Knowledge (Attacks Scenarios)}

In this work, we consider three attack scenarios:

\noindent
    \textbf{\textit{Transferability Attack.}} We assume the adversary is aware of the exact classifier internal model; its architecture and parameters. The adversary uses the exact classifier to create adversarial examples. Thus, we explore whether these examples transfer effectively to the approximate classifier (DA classifier).
    
  \noindent  
    \textbf{\textit{Black-box Attack.}} We assume the attacker has access only to the input/output of the victim classifier (which is our approximate classifier) and has no information about its internal architecture.  The adversary first uses the results of querying the victim to reverse engineer the classifier and create a substitute 
    CNN model.   With the substitute model, the attacker can attempt to generate different adversarial examples to attack the victim classifier.

\noindent
    \textbf{\textit{White-box Attack.}} We assume a powerful attacker who has full knowledge of the victim classifier's architecture and parameters (including the fact that it uses AC).  The attacker uses this knowledge to create adversarial examples.  

\subsection{Adversarial Example Generation}

 We consider several adversarial attack generation algorithms for our attack scenarios, including some of the most recent and potent evasion attacks.  Generally, in each algorithm, the attacker tries to evade the system by adjusting malicious samples during the inference phase, assuming no influence over the training data. However, as different defenses have started to be deployed that specifically target individual adversarial attack generation strategies, new algorithms have started to be deployed that bypass these defenses. For example, methods such as defensive distillation \cite{distillation_SP} and automation detection \cite{early} were introduced and demonstrate robustness against the Fast gradient sign attack 
 \cite{fgsm}.  However, the new $C\&W$ attack was able to bypass these defenses \cite{CW}.  
Thus, demonstrating robustness against a range of these attacks provides confidence that a defense is effective in general, against all known attack strategies, rather than against a specific strategy.

These attacks can be divided into three categories: Gradient-based attacks relying on detailed model information, including the gradient of the loss w.r.t. the input. Score-based attacks rely on the predicted scores, such as class probabilities or \textit{logits} of the model. On a conceptual level, these attacks use the predictions to estimate the gradient numerically.  Finally, decision-based attacks rely only on the final output of the model and minimizing the adversarial examples' norm.  The attacks are summarized in Table \ref{Attack_methods}. 

\begin{table}[!htp]
\small
\centering
  \caption{Summary of the used attack methods. Notice that the strength estimation is based on~\cite{strength}.} 
  \label{Attack_methods}
  \begin{tabular}{ccccc}
    \toprule
    \textbf{Method} & \textbf{Category}  & \textbf{Norm} & \textbf{Learning} & \textbf{Strength} \\
    \midrule
    FGSM~\cite{fgsm}  & gradient-based   &  $L_\infty$  & One shot  & ***\\
    PGD~\cite{pgd}   & gradient-based   &  $L_\infty$  & Iterative & ****\\
    JSMA~\cite{SMA}  & gradient-based   &  $L_0$       & Iterative & ***\\
    C\&W~\cite{CW}  & gradient-based   &  $L_2$       & Iterative & *****\\
    DF~\cite{deepfool}    & gradient-based   &  $L_2$       & Iterative & ****\\
    LSA~\cite{localsearch}   & Score-based      &  $L_2$       & Iterative & ***\\
    BA~\cite{BA}    & Decision-based   &  $L_2$       & Iterative & ***\\
    HSJ~\cite{HSJ}    & Decision-based   &  $L_2$       & Iterative & *****\\
    
  \bottomrule
\end{tabular}
\end{table}

\label{sec:threat}

\section{Defensive Approximation: Implementing Approximate CNNs}

We propose to leverage approximate computing to improve the robustness of ML classifiers, such as CNNs, against adversarial attacks. We call this general approach \textbf{Defensive Approximation} (DA). The closest approach to DA is the perturbation-based defense~\cite{snP2019_certif,smooth} that either adds noise or otherwise filter the input data to try to interfere with any adversarial modifications to the input of a classifier. However, our approach advances the state-of-the-art by injecting perturbations throughout the classifier and directly by the approximate hardware, thereby enhancing both robustness and power efficiency. 

Technically, the approximate design process is driven by two main considerations:
\begin{enumerate}[label=(\alph*)]
    \item Injecting significant noise that can influence the CNN behavior, and allows by-product power gains. 
    \item Keeping the cumulative noise magnitude under control to avoid impacting the baseline accuracy of the CNN. 
\end{enumerate}

For consideration (b), we purpose to an implementation that replaces the conventional mantissa multiplier in floating point multipliers, with a simpler approximate design. This choice is backed by the study in \cite{iccd18, D&T}, which show that errors in the exponent part might have a drastic impact on CNNs accuracy. 
Consideration (a) means that the approximate design needs to be aggressive to induce significant noise. Therefore, as detailed in the next subsection, we chose the corner case, i.e., the most aggressive approximate design \cite{AMA5, Heap}, and used it to replace the mantissa computation logic in a basic floating point multiplier. 

In this section, we present our approximate multiplier design and analyze its properties.

\subsection{Approximate Floating Point Multiplier}
ML structures, such as CNNs, often rely on computationally expensive operations, e.g., convolutions that are composed of multiplications and additions. Floating-point multiplications consume most of the processing energy in both inference and training of CNNs~\cite{esl_ihsen,Heap}. 
Although approximate computation can be introduced in different ways (with likely different robustness benefits), DA leverages a new approximate 32-bit floating-point multiplier, which we call \textit{approximate floating-point multiplier} (Ax-FPM).
The IEEE 754-2008 compliant floating-point format binary numbers are composed of three parts: a sign, an exponent, and a mantissa (also called fraction)~\cite{FP}. 
The sign is the most significant bit, indicating whether the number is positive or negative. In a single-precision format, the following $8$ bits represent the exponent of the binary number ranging from $-126$ to $127$. The remaining $23$ bits represent the fractional part (mantissa). 
For most of the floating number range, the normalized format is:
\begin{equation}\label{eqn:ieee754}
val = ( -1 )^{\textit{sign}} \times 2^{exp-bias} \times ( 1.fraction ) 
\end{equation}

A floating-point multiplier (FPM) consists mainly of three units: mantissa multiplier, exponent adder, and a rounding unit.  The mantissa multiplication consumes $81\%$ of the overall power of the multiplier \cite{trunca}.

Ax-FPM is designed based on a mantissa multiplication unit that is constructed using approximate full adders (FA). The FAs are aggressively approximated to inject computational noise within the circuit. We describe Ax-FPM by first presenting the approximate FA design, and then the approximate mantissa multiplier used to build the Ax-FPM. 



\begin{figure}[!htp]
\centering
\includegraphics[width=\columnwidth]{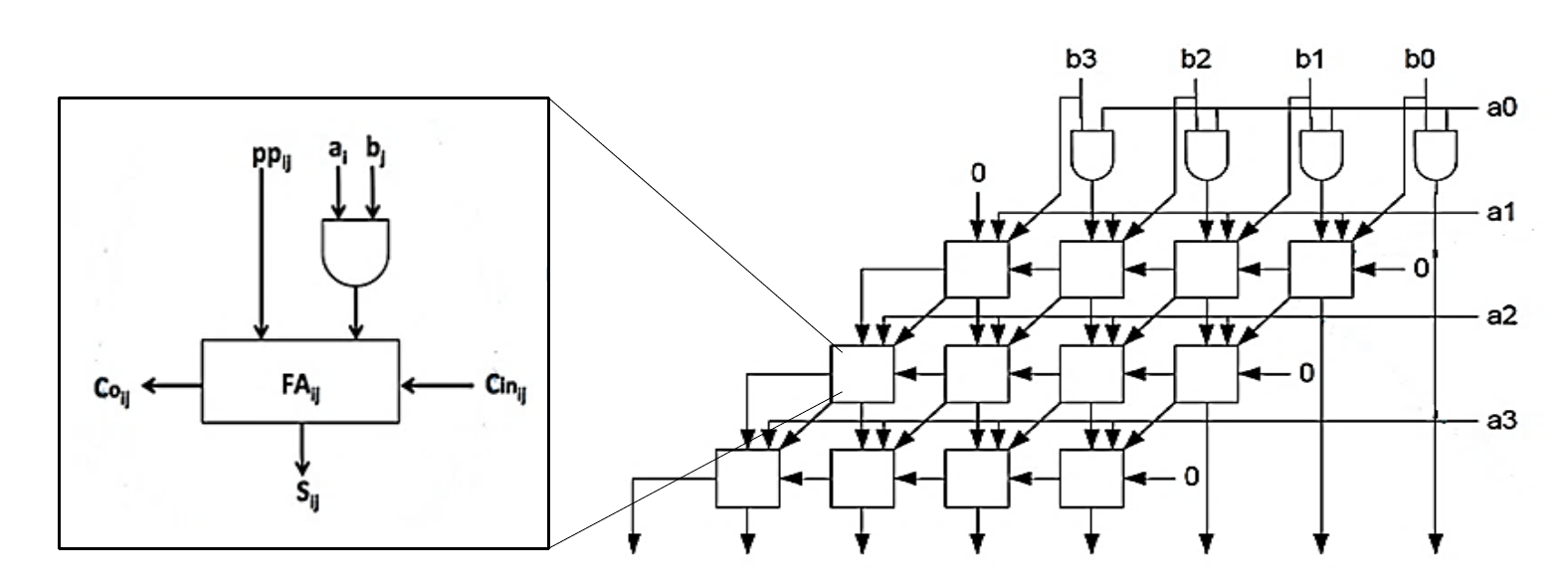}
\caption{An illustration of a $4 \times 4$ array multiplier architecture.}
\label{MA}
\end{figure}

To build a power-efficient and a higher performance FPM, we propose to replace the mantissa multiplier by an approximate mantissa multiplier; an array multiplier constructed using approximate FAs. We selected an array multiplier implementation because it is considered one of the most power-efficient among conventional multiplier architectures~\cite{array}.  In the array architecture, multiplication is implemented through the addition of partial products generated by multiplying the multiplicand with each bit of multiplier using AND gates, as shown in Figure \ref{MA}.  
\begin{figure}[!htp]
\centering
\includegraphics[width=\columnwidth]{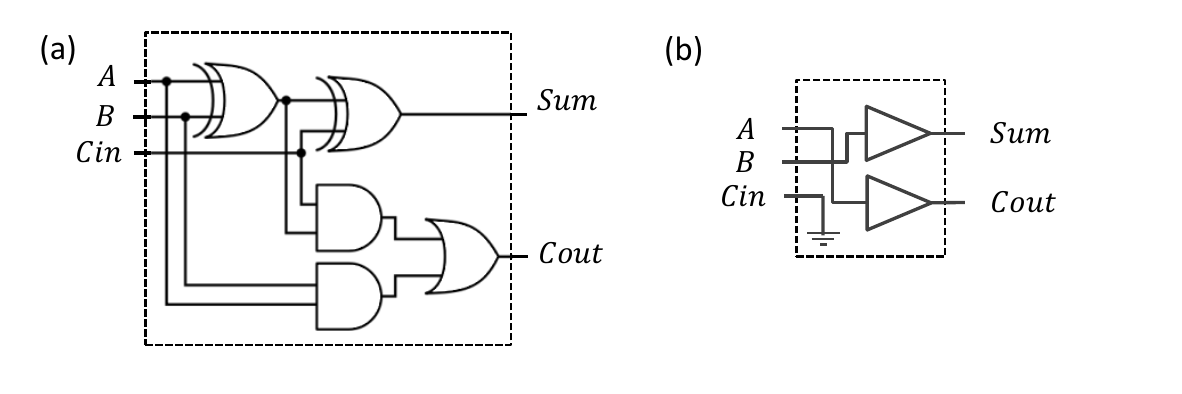}
\caption{Logic diagram of (a) exact Full Adder, (b) AMA5 (used in this work).}
\label{FA}
\end{figure}

Specifically, we build an array multiplier based on an approximate mirror adder (AMA5)~\cite{AMA5} in place of exact FAs. The approximation of a conventional FA is performed by removing some internal circuitry, thereby resulting in power and resource reduction at the cost of introducing errors.  Consider a FA with (A,B, $C_{in}$) as inputs and ($Sum$, $C_{out}$) as outputs ($C$ here refers to carry). For any input combination, the logical \textit{approximate} expressions for $Sum$ and $C_{out}$ are: $Sum = B$ and $C_{out} = A$. 
The AMA5 design is constituted by only two buffers (see Figure \ref{FA}), leading to the latency and energy savings relative to the exact design, but more importantly, introduce errors into the computation.  It is worth noting that these errors are data dependent, appearing for specific combinations of the inputs, and ignoring the carry in value, making the injected noise difficult to predict or model.  

When trying to evaluate the proposed Ax-FPM, we were interested in studying its behavior under small input numbers ranging between $-1$ and $+1$ since most of the internal operations within CNNs are in this range. 

\begin{figure}[!htp]
\centering
\includegraphics[width=0.8\columnwidth]{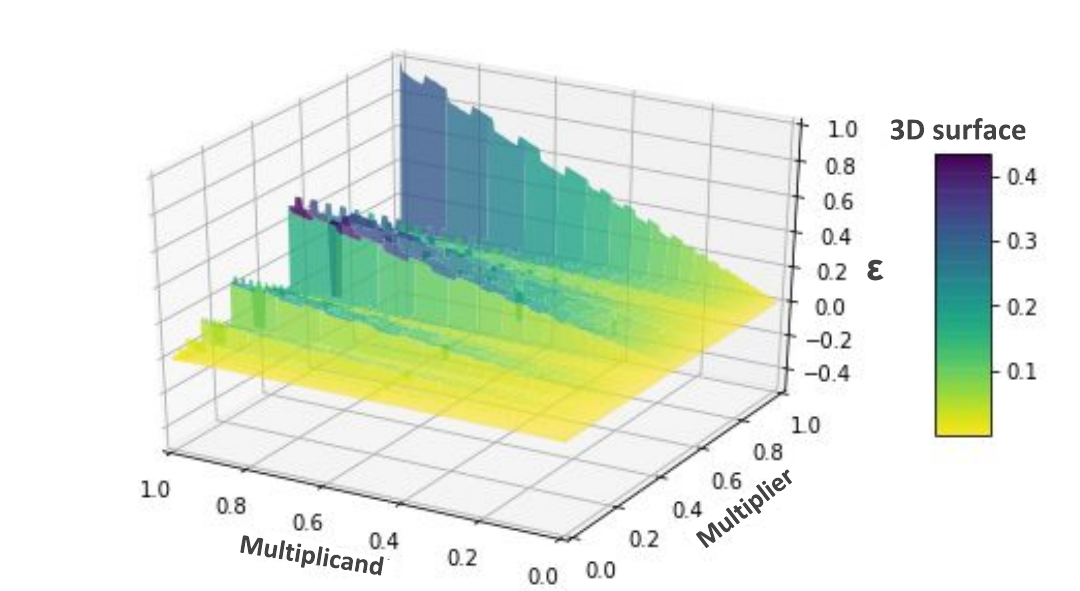}
\caption{Noise introduced by the approximate multiplier while the operands in $[0,1]$.
}
\label{appx_mult}
\end{figure}

We measure the introduced error as the difference of the output of the approximate multiplier and the exact multiplier. The results are shown in Figure \ref{appx_mult} using 100 million randomly generated multiplications across the input range from $-1$ to $1$. 
Three trends can be observed that will be used later to help understanding the impact of the approximation using Ax-FPM on CNN security: \textbf{(i)} The first is the data-dependent discontinuity of the approximation-induced errors, \textbf{(ii)}  We noticed that in $96\%$ of the cases, the approximate multiplication results in higher absolute values than the exact multiplication: For positive products, the approximate result is higher than the exact result, and for negative product results the approximate result is lower than the exact result, and \textbf{(iii)} In general, we notice that the larger the multiplied numbers, the larger the error magnitude added to the approximate result. As shown later, these observations will help understand the mechanism that we think is behind robustness enhancement. 

\begin{figure}[!htp]
\centering
\includegraphics[width=0.9\columnwidth,height=4cm]{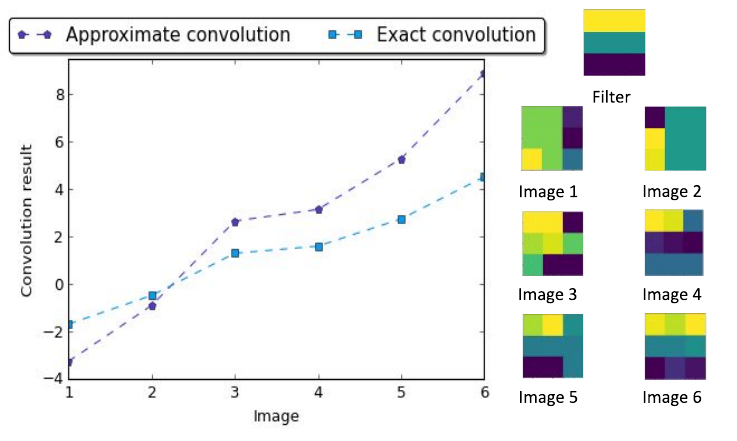}
\caption{Convolution result of the filter and each input image using exact and approximate convolution.}
\label{appx_conv}
\end{figure}

\subsection{Approximate Convolution}

In order to understand the impact of AC at larger scales than the individual multiplication, we track the impact on convolution operations. The approximate CNN is built using the approximate convolution operations as building blocks. The activation functions and the pooling layers which do not use multiplication are similar to the conventional CNN.   Convolution layers enable CNNs to extract data-driven features rather than relying on manual feature-extraction in classical ML systems. The convolution operation is performed on the layer's input data using a kernel matrix that is convoluted (piece-wise multiplied) against the input data to produce a feature map. 



As we slide the filter over the input from left to right and top to bottom whenever the filter coincides with a \textbf{similar} portion of the input, the convolution result is high, and the neuron will fire. 
The weight matrix filters out portions of the input image that does not align with the filter, and the approximate multiplier helps improve this process by further increasing the output when a feature is detected. In Figure \ref{appx_conv}, we run an experiment where we choose a filter and six different images with different degrees of similarity to the chosen filter (1 to 6 from the least to the most similar), and we perform the convolution operation. We notice that the approximate convolution using Ax-FPM delivers higher results for similar inputs and lower results for dissimilar inputs. We can also notice that the higher the similarity, the higher the gap between the exact and Ax-FPM approximation results.  Therefore, by using the approximate convolution, the main features of the image that are important in the image recognition are retained and further highlighted with higher scores that will later help increase the confidence of the classification result, as explained in Section \ref{sec:how}.

\subsection{AC Design Space Exploration}
For the sake of comparison, we proceed to a design space exploration considering the accuracy of the multipliers and resource utilization as optimization objectives. The comparison of our design with the optimal approximate design given by the exploration and referred to as HEAP led to a clearly dominant choice in favor of our design given the following perspectives:

\textbf{(i) Resource and power efficiency:} Given its aggressive design, Ax-FPM achieves the lowest power consumption and resource utilization. 

\textbf{(ii) Robustness:} While other approximate designs had a positive impact on CNNs robustness, we noticed that Ax-FPM achieves the highest enhancement. 

\textbf{(iii) Accuracy:} CNNs are highly approximation-tolerant, and even aggressive multiplier design didn't impact overall accuracy. Ax-FPM, as well as HEAP have insignificant impact on accuracy. 

While these observations give an insight on the potential generalizability of AC positive impact, Ax-FPM is a non dominated design outperforming other design in both robustness, power consumption and resource utilization, while having the same overall accuracy level. 
Hence, in the remainder of the paper, we will focus only on Ax-FPM based DA. Further details and results can be found in the Appendix.

\label{sec:AppxCNN}

\section{Can DA help in defending against adversarial attacks? }

In this section, we empirically explore the robustness properties of DA under a number of threat models.  
We first explore the transferability of adversarial attacks where we evaluate whether attacks crafted for exact CNNs transfer to approximate CNNs. We then consider direct attacks against approximate CNNs in both black and white-box settings.  


\subsection{Experimental Setup}

The first benchmark we use is the LeNet-5 CNN architecture \cite{lenet5} along with the MNIST database \cite{mnist}, which implements a classifier for handwritten digit recognition. The MNIST consists of 60,000 training and 10,000 test images with 10 classes corresponding to digits. Each digit example is represented as a gray-scale image of $28 \times 28$ pixels, where each feature corresponds to a pixel intensity normalized between 0 and 1.
We also use the AlexNet image classification CNN \cite{Alexnet} along with the CIFAR-10 database \cite{CIFAR}.  CIFAR-10 consists of 60,000 images, of dimension $64 \times 64 \times 3$ each and contains ten different classes.  LeNet-5 consists of two convolutional layers, two max-pooling layers, and two fully connected layers.  AlexNet uses five convolution layers, three max-pooling layers, and three fully connected layers. 
The rectified linear unit (ReLU) was used as the activation function in this evaluation, along with a dropout layer to prevent overfitting.
For both models, the output layer is a special activation function called softmax that will assign probabilities to each class. 
Our implementations are built using the open source ML framework PyTorch \cite{PyTorch}.
We use the Adam optimization algorithm to train the LeNet-5 classifier.  For Alexnet, we use Stochastic Gradient Descent (SGD) with a learning rate equal to  $0.01$ and $0.001$, respectively.  Note that \emph{no retraining} is applied in DA, we rather use the same hyper-parameters obtained from the original (exact) classifier; we simply replace the multipliers with the approximate multiplier.

Our reference exact CNNs are conventional CNNs based on exact convolution layers with the format Conv2d provided by PyTorch.  In contrast, the approximate CNNs emulate the 32-bit Ax-FPM functionality and replace the multiplication in the convolution layers with Ax-FPM in order to assess the behavior of the approximate classifier. 

Since we are simulating a cross-layer (approximate) implementation from gate-level up to system-level, the experiments (forward and backward gradient) are highly time consuming, which limited our experiments to the two datasets MNIST and CIFAR-10.

Except for the black-box setting where the attacker trains his own reverse-engineered proxy/substitute model, the approximate and exact classifiers share the same pre-trained parameters and the same architecture; they differ only in the hardware implementation of the multiplier.


\subsection{Do Adversarial Attacks on an Exact CNN Transfer to an Approximate CNN ?}
\label{sec:transExApx}

\begin{figure}[!htp]
\centering
\includegraphics[width=\columnwidth]{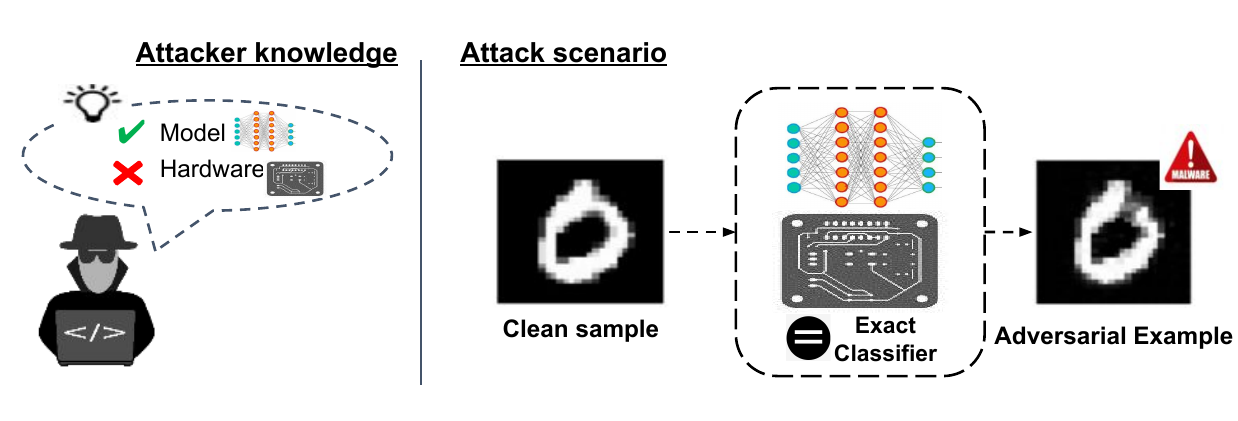}
\caption{Transferrability attack scenario.} 
\label{grey-box}
\end{figure}

\textbf{Attack Scenario.} In this setting, the attacker has full knowledge of the classifier architecture and hyper-parameters, but without knowing that the model uses approximate hardware.  An example of such scenario could be in case an attacker correctly guesses the used architecture based on its widespread use for a given application (e.g., LeNet-5 in digit recognition), but is unaware of the use of DA as illustrated in Figure~\ref{grey-box}.

\noindent\textbf{Transferability Analysis.} The classifier generates adversarial examples using the set of algorithms in Table~\ref{Attack_methods} and assume that the exact classifier from Lenet-5 trained on the MNIST dataset. Notice that the hyperparameters, as well as the structure of the network, are the same between the exact and the approximate classifiers.  The attacker then tests the adversarial examples against the approximate classifier.  
Table \ref{grey-boxresult} presents the attacks respective success rates. We notice that the DA considerably reduces the transferability of the malicious samples and, by consequence, increases the robustness of the classifier to this attack setting. We observed that the robustness against transferability is very high, and reaches $99\%$ for $C\&W$ attack.


\begin{table}[!htp]
\small
  \caption{Attacks transferability success rates for MNIST.}
  \label{grey-boxresult}
  \begin{tabular}{ccc}
    \toprule
    \textbf{Attack method} & \textbf{Exact LeNet-5}  & \textbf{Approximate LeNet-5} \\
    \midrule
        FGSM          &   100\%    & 12\%    \\
        PGD           &   100\%    & 28\%    \\
        JSMA          &   100\%    & 9\%   \\
        C\&W          &   100\%    & 1\%    \\
        DF	          &   100\%    & 17\%   \\ 
        LSA	          &   100\%    & 18\%   \\ 
        BA	          &   100\%    & 17\%   \\ 
        HSJ	          &   100\%    & 2\%    \\         
  \bottomrule
\end{tabular}
\end{table}

We repeat the experiment for AlexNet with CIFAR-10 dataset. For the same setting, the success of different adversarial attacks is shown in Table \ref{attackAlexnet}.  While more examples succeed against the approximate classifier, we see that the majority of the attacks do not transfer.  Thus, DA offers built-in robustness against transferability attacks.  

\begin{table}[!htp]
\small
\centering
  \caption{Attacks transferability success rates for CIFAR-10.}
  \label{attackAlexnet}
  \begin{tabular}{ccc}
    \toprule
    \textbf{Attack method} & \textbf{Exact AlexNet}  & \textbf{Approximate AlexNet} \\
    \midrule

        FGSM          &   100\%    & 38\%    \\        
        PGD           &   100\%    & 31\%    \\
        JSMA          &   100\%    & 32\%   \\        
        C\&W         &    100\%  & 17\%    \\
        DF	          &   100\%    & 35\%   \\ 
        LSA	          &   100\%    & 36\%   \\ 
        BA	          &   100\%    & 37\%   \\ 
        HSJ	          &   100\%    & 12\%    \\ 
  \bottomrule
\end{tabular}
\end{table}

Notice that, unlike other state-of-the-art defenses, our defense mechanism protects the network without relying on the attack details or the model specification and without any training beyond that of the original classifier.  
Unlike most of the perturbation-based defenses that degrade the classifier’s accuracy on non-adversarial inputs, our defense strategy significantly improves the classification robustness with no baseline accuracy degradation, as we will show in Section~\ref{CNNaccuracy}.  


\subsection{Can We Attack an Approximate CNN?}

In the remaining attack models, we assume that the attacker has direct access to the approximate CNN. We consider both black-box and white-box settings.

\begin{figure}[!htp]
\centering
\includegraphics[width=\columnwidth]{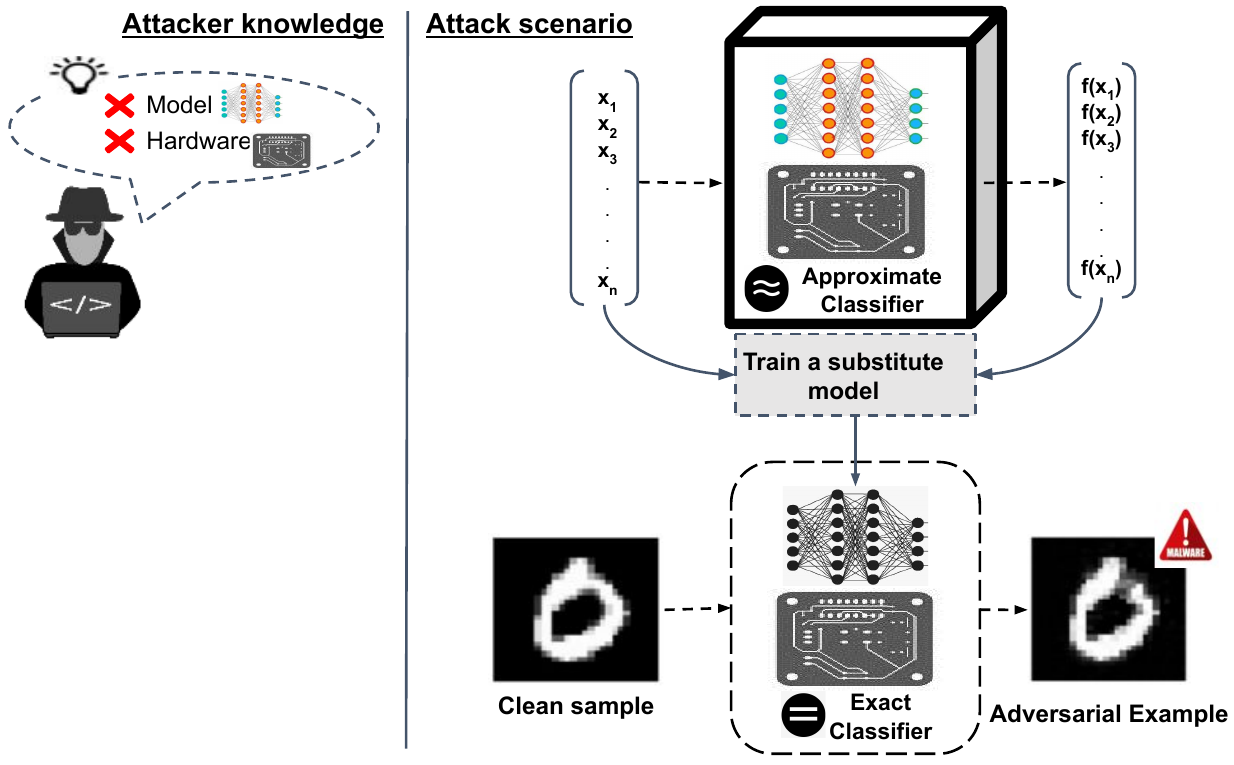}
\caption{Black-box attack scenario. }
\label{black-box}
\end{figure}

\textbf{Black-box Attack.} In a black-box setting, the attacker has no access to the classifier architecture, parameters, and the hardware platform but can query the classifier with any input and obtain its output label.  In a typical black-box attack, the adversary uses the results of many queries to the target model to reverse engineer it.  Specifically, the adversary trains a substitute (or proxy) using the labeled inputs obtained from querying the original model (see Figure \ref{black-box}). 
We also conduct a black box attack on the exact classifier and evaluate how successful the black box attack is in fooling it.  Essentially, we are comparing the black-box transferability of the reverse-engineered models to the original models for both the exact and the approximate CNNs. 

In Table \ref{black-boxresult}, we present the attack success ratios for the exact CNN and the approximate/DA CNN. DA increases resilience to adversarial attacks across various attacks and for both single-step and iterative ones: it achieves $73\%$ classification success on adversarial examples in the worst case and the defense succeeded in up to $100\%$ of the examples generated by C\&W, PGD, and HSJ respectively.  
\begin{table}[!htp]
\small
\centering
  \caption{Black-box attacks success rates for MNIST.}
  \label{black-boxresult}
  \begin{tabular}{ccc}
    \toprule
    \textbf{Attack method} & \textbf{Exact LeNet-5}  & \textbf{Approximate LeNet-5} \\
    \midrule
        FGSM          &   100\%    & 22\%    \\
        PGD           &   100\%    & 0\%    \\
        JSMA          &   100\%    & 13\%   \\
        C\&W          &    100\%   & 0\%    \\
        DF	          &   100\%    & 25\%   \\ 
        LSA	          &   100\%    & 26\%   \\ 
        BA	          &   100\%    & 27\%   \\ 
        HSJ	          &   100\%    & 0\%    \\ 
  \bottomrule
\end{tabular}
\end{table}

\begin{figure}[!htp]
\centering
\includegraphics[width=\columnwidth]{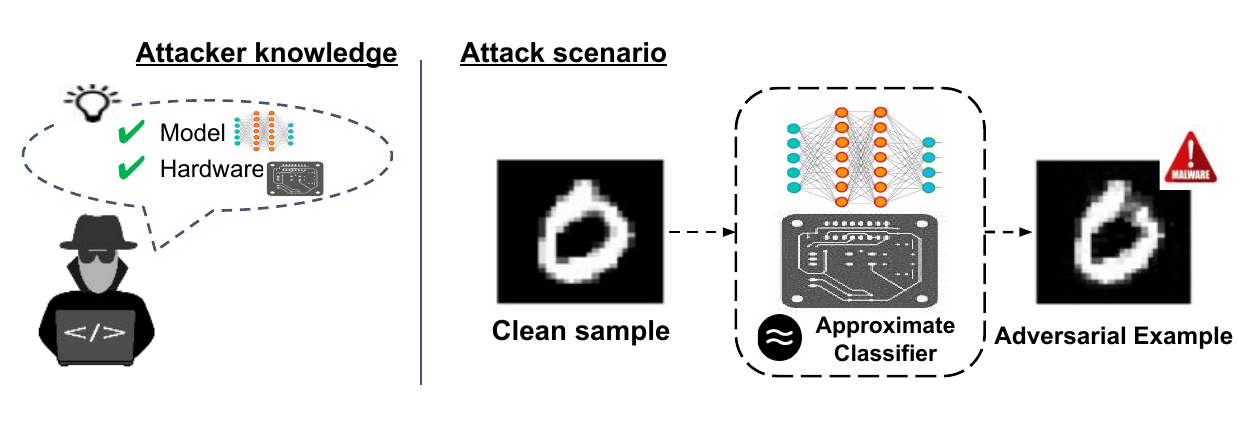}
\caption{White-box attack scenario.}
\label{white-box}
\end{figure}

\textbf{White-box Attack.} In this setting, the attacker has access to the approximate hardware along with the victim model architecture and parameters, as shown in Figure \ref{white-box}. In particular, the adversary has full knowledge of the defender model, its architecture, the defense mechanism, along with full access to approximate gradients used to build the gradient-based attacks. In essence, the attacker is aware of our defense, and can adapt around it with full knowledge of the model and the hardware, which is a recommended methodology for evaluating new defenses~\cite{carlini_gift}.


In this scenario, we assume a powerful attacker with full access to the approximate classifier's internal model and can query it indefinitely to directly create adversarial attacks.  Although DA in production would normally reduce execution time, in our experiments, we \emph{emulate} the 32-bit Ax-FPM functionality within the approximate classifier.  As a result, this makes inference extremely slow: on average, it takes $5$ to $6$ days to craft one adversarial example on an 8th Gen Intel core i7-8750H processor with NVIDIA GeForce GTX 1050. This led us to limit the white-box experiments; we use only two of the most efficient attacks in our benchmark: $C\&W$ and DeepFool attacks, and for a limited number of examples selected randomly from our test set for different classes. 

In a white box attack, with an unconstrained noise budget, an adversary can always eventually succeed in causing an image to misclassify. Thus, robustness against this type of attack occurs through the magnitude of the adversarial noise to be added: if this magnitude is high, this may exceed the ability of the attacker to interfere, or cause the attack to be easily detectable. 
\begin{figure}[!htp]
\centering
\includegraphics[width=\columnwidth]{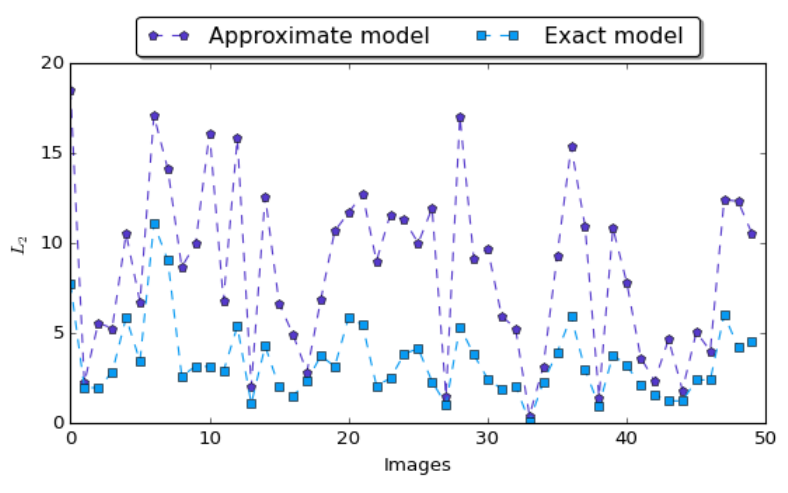}
\caption{Required perturbations by DeepFool attack, measured using $L_2$ distance, to generate MNIST adversarial examples for approximate and exact classifiers. }
\label{L2}
\end{figure}

\begin{figure}[!htp]
\centering
\includegraphics[width=0.9\columnwidth]{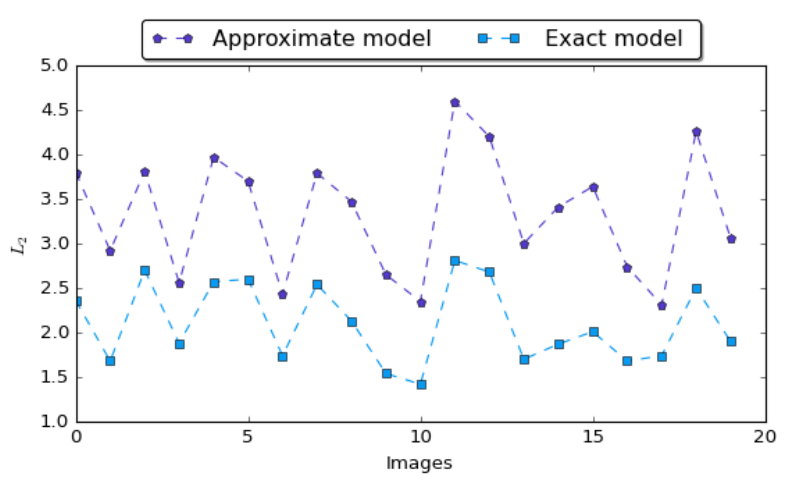}
\caption{Required perturbations by C\&W attack, measured using $L_2$ distance, to generate MNIST adversarial examples for approximate and exact classifiers. }
\label{L2_C&W}
\end{figure}

Figures \ref{L2} and \ref{L2_C&W}, respectively, present different measures of $L_2$ for adversarial examples crafted using DF and $C\&W$ attacking both a conventional CNN and an approximate CNN. We notice that the distance between a clean image and the adversarial example generated under DA is much larger than the distance between a clean sample and the adversarial example generated for the exact classifier. On average, a difference of $5.12$ for $L_2$-DeepFool attacks and $1.23$ for $L_2$-C\&W attack. 
This observation confirms that DA is more robust to adversarial perturbations since the magnitude of the adversarial noise has to be significantly higher for DF to fool DA successfully.  

\begin{figure}[!htp]
\centering
\includegraphics[width=\columnwidth]{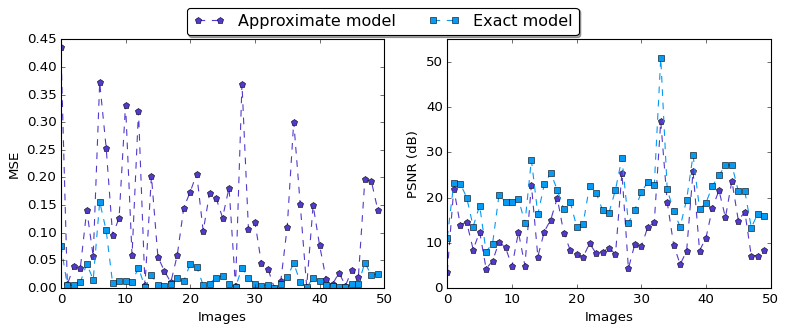}
\caption{MSE and PSNR values for the generated adversarial examples using DeepFool method when attacking the approximate and the exact classifiers.}
\label{MSE&PSNR_DF}
\end{figure}

\begin{figure}[!htp]
\centering
\includegraphics[width=\columnwidth]{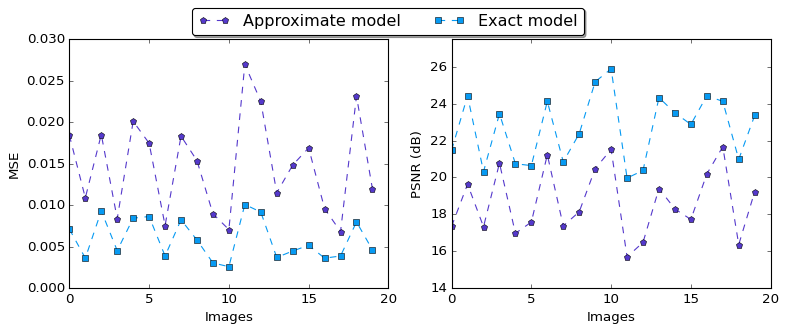}
\caption{MSE and PSNR values for the generated adversarial examples using $L_2$-C\&W method when attacking the approximate and the exact classifiers.}
\label{MSE&PSNR}
\end{figure}

To understand the implication of this higher robustness in terms of observable effects on the input image,  we also show the Peak Signal to Noise Ratio (PSNR) and the Mean Square Error (MSE) in Figures \ref{MSE&PSNR_DF} and \ref{MSE&PSNR}; these are two common measures of the error introduced in a reconstruction of an image. Specifically,  MSE represents the average of the squares of the "errors" between the clean image and the adversarial image. The error is the amount by which the values of the original image differ from the distorted image. PSNR is an expression for the ratio between the maximum possible value (power) of a signal and the power of distorting noise that affects the quality of its representation. It is given by the following equation: $P S N R=20 \log _{10}\left(\frac{M A X_{x}}{\sqrt{M S E}}\right)$. The lower the PSNR, the higher the image quality degradation is.

We notice that the adversarial examples generated for DA is more noisy than adversarial examples generated for an exact classifier. The PSNR difference reaches $4 dB$ for C\&W and $7.8dB$ for DeepFool. Moreover, on average, the DA-dedicated adversarial examples have $6$ times, and $3$ times more MSE than the exact classifier-dedicated adversarial examples for $C\&W$ and DeepFool attacks, respectively. 


We can conclude that DA provides substantial built-in robustness for all three attack models we considered.  Attacks generated against an exact model do not transfer successfully to DA.  Black-box attacks also achieve a low success rate against DA.  Finally, even white-box attacks require substantial increases in the injected noise to fool DA.  Next, we probe deeper into DA's internal behavior to provide some intuition and explanation for these observed robustness advantages.

\label{sec:secu}

\section{How does DA help CNN robustness?}
In this section, we probe into the DA classifier's operation to attempt to explain the robustness advantages we observed empirically in the previous section.  
While the explainability of deep neural networks models is a known hard problem, especially under adversarial settings~\cite{samek2019explainable}, we attempt to provide an overview of the mechanisms that we think are behind the DA impact on security. 
We study the impact of the approximation on CNNs' confidence and generalization property. We follow this analysis in the Appendix with a mathematical argument explaining the observed robustness based on recent formulations by Lecuyer et al.~\cite{snP2019_certif}. 


The output of the CNN is computed using the \emph{softmax} function, which normalizes the outputs from the fully connected layer into a likelihood value for each output class.  Specifically, this function takes an input vector and returns a non-negative probability distribution vector of the same dimension corresponding to the output classes.   In this section, we examine the impact of approximation on the observed classifier confidence. We compare the output scores of an exact and an approximate classifier for a set of $1000$ representative samples selected from the MNIST dataset: $100$ randomly selected from each class. We define the classification confidence, $C$, as the difference between the true class $l$'s score and the "runner-up" class score, i.e., the class with the second-highest score. $C$ is expressed by $C = output[l] - max_{j \neq l} \{output[j]\}$.  The confidence ranges from $0$ when the classifier gives equal likelihood to the top two or more classes, to $1$ when the top class has a likelihood of $1$, and all other classes $0$.



\begin{figure}[!htp]
\centering
\includegraphics[width=0.9\columnwidth]{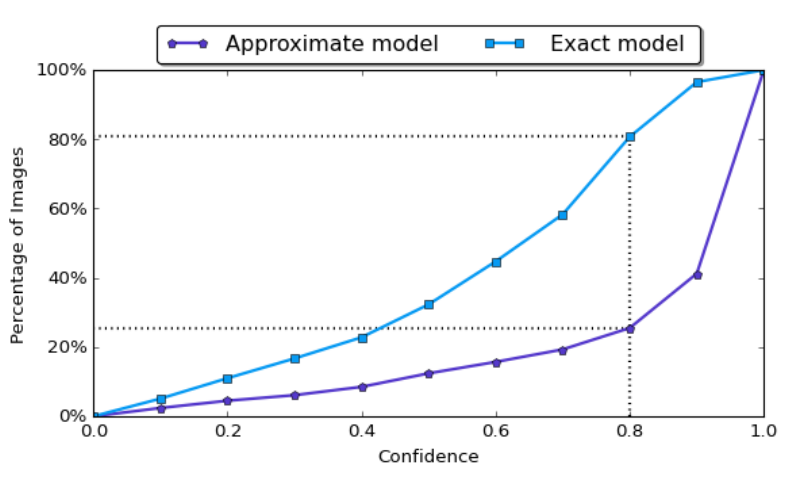}
\caption{Cumulative distribution of confidence.}
\label{conf_dis}
\end{figure}

We plot the cumulative distribution of confidence for both classifiers in Figure \ref{conf_dis}.  DA images have higher confidence; for example, in images classified by the exact classifier, less than $20\%$  had higher than $0.8$ confidence.   
On the other hand, for the approximate classifier, $74.5\%$ of the images reached that threshold.


Compared to the baseline feature maps generated from an exact convolution operation, for the same pre-trained weights, our approximate convolution highlights further the features. Recall that the multiplier injected noise is higher when input numbers are higher (i.e., there is a high similarity between the kernel and the input data) and lower when the inputs are lower (when the similarity is small), as shown in Figure \ref{appx_conv}. We believe that these enhanced features continue to propagate through the model resulting in a higher probability for the predicted class. This higher confidence requires thereby higher noise to decrease the true label's likelihood, and increase another label's.  

\label{sec:how}

\section{How does DA compare to other reduced precision techniques?}

In this section, we investigate the impact of other reduced precision techniques on robustness. We first compare DA to DQ, and then study the impact of using Bfloat16 data representation on the system performance and robustness.

\subsection{Defensive Quantization}

Defensive quantization \cite{DQ} was proposed to jointly optimize the efficiency and robustness of deep learning models. A 4-bit quantized model was trained on CIFAR-10 using the Dorefa-Net method \cite{zhou2018dorefanet}. The model architecture is detailed in Appendix \ref{mod_archi}. We consider two ways of quantization: (i) Weight quantization, where only the weights are quantized, and (ii) Full quantization where the weights 
of each convolutional and dense block and the output of each activation function are quantized. 
In Table \ref{quan_result4}, we report transferability between exact (32-bit floating point), approximate model (using DA), fully quantized and weight-only quantized model.
We notice that DA is almost  two times more robust against transferability attacks than DQ under FGSM, PGD and C\&W attacks. We believe that this is due to the difference in terms of noise distribution between DA and DQ. In fact, while DQ-induced noise tends to make the initial decision boundary smoother, the input-dependent noise in DA makes its decision boundary randomly different from the initial model.

\begin{table}[!htp]
\small
  \caption{Comparing attacks transferability success rates for CIFAR-10 when using DA and DQ.}
  \label{quan_result4}
  \begin{tabular}{ccccc}
    \toprule
                     &          & \textbf{DA:} & \textbf{DQ:} & \textbf{DQ:} \\
    \textbf{Attack method} & \textbf{Exact}  &  & \textbf{Full } & \textbf{Weight-only } \\
    \midrule
        FGSM          &   100\%    & 38\%  & 60\%  & 61\% \\
        PGD           &   100\%    & 31\%  & 74\% & 73\%\\
        C\&W          &   100\%    & 17\%  & 68\% &  68\% \\
        
  \bottomrule
\end{tabular}
\end{table}

\subsection{BFloat16}

The Bfloat16 (Brain Floating Point) \cite{ kalamkar2019study} is a truncated version of the 32-bit IEEE 754  single-precision floating-point format (float32). It is composed of one sign bit, eight exponent bits, and seven mantissa bits giving the range of a full 32-bit number but in the data size of a 16-bit number. Bfloat16 is used in machine learning applications and intended to reduce the storage requirements and increase computing speed without losing precision. 

To evaluate the impact of Bfloat16 on deep neural networks robustness, we use Pytorch framework \cite{PyTorch} to implement Bfloat16-based CNN architectures and test them for MNIST and CIFAR-10 benchmarks.
We notice that using Bfloat16 achieves the same prediction accuracy as the full precision 32-bit floating point. No remarkable change was noticed at the output of the convolutional layers nor in the confidence of the models. We believe this is due to the nature of the noise introduced by reducing data representation accuracy. A more detailed discussion can be found in Section \ref{discussion}. 

\begin{figure}[!htp]
\centering
\includegraphics[width=0.9\columnwidth]{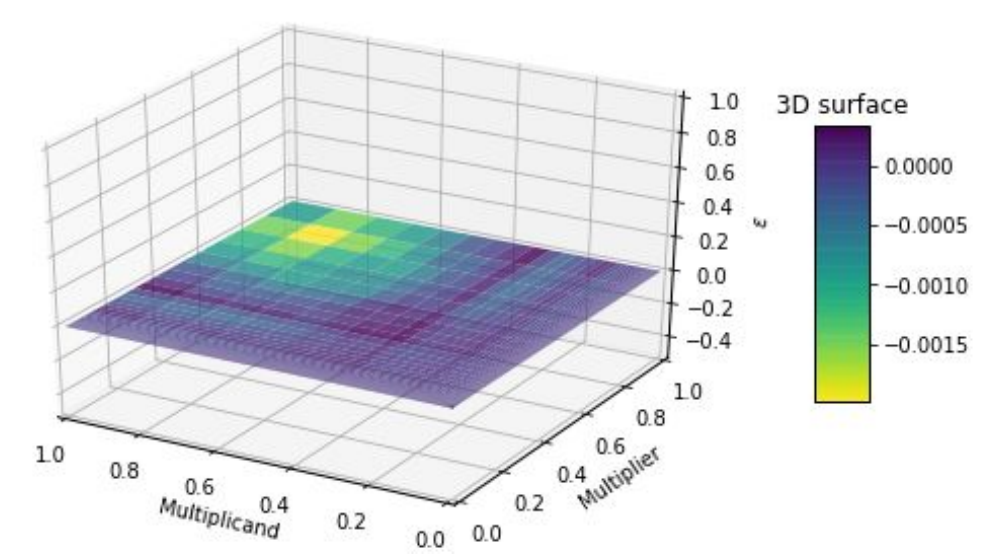}
\caption{Noise introduced by multiplying Bfloat16 while the operands $\in [0,1]$.
}
\label{bfloat16}
\end{figure}

Figure \ref{bfloat16} shows the noise resulting from multiplying $100$ million randomly generated Bfloat16 numbers compared to their corresponding 32-bit floating point. This figure shows that, in contrast with our approximate multiplier (Figure \ref{appx_mult}), Bfloat16 multiplication results in mostly negative noise with orders of magnitude lower than DA-induced noise. Moreover, the Bfloat16 noise is not input-dependent, and has no specific impact on the model confidence. Accordingly, no improvement in the robustness was noticed when attacking the Bfloat16 model using FGSM, PGD and $C\&W$. 

\label{sec:quan}

\section{Baseline Performance Implications}
\subsection{Impact on Model Accuracy} 
\label{CNNaccuracy}
It is important for a defense mechanism that aims to enhance robustness against adversarial attacks to keep at least an acceptable performance level for clean inputs. In fact, considerably reducing the baseline accuracy, or creating an exploding or vanishing gradient impact that makes the model sensitive to other types of noise undermines the model reliability.  
In our proposed approach, we maintain the same level of recognition rate even with the approximate noise in the calculations. Counter-intuitively, this data-dependent noise helps to better highlight the input's important features used in the recognition and does not affect the classification process. 
A drop of $0.01 \%$ in the recognition rate for the case of LeNet-5 and $1\%$ for AlexNet is recorded as mentioned in Table \ref{tab3}. 


\begin{table}[!htp]
\small
\centering
  \caption{Accuracy results of Float32, approximate model, fully quantized, weight-only quantized and Bfloat16 models.}
  \label{tab3}
  \begin{tabular}{ccc}
    \toprule
    \textbf{Used Multiplier }& \textbf{MNIST} & \textbf{CIFAR-10}\\
    \midrule
        Float32                 & 97.93\%  &   81\%   \\
        Approximate (DA)	    & 97.67\%  &   80\%     \\ 
        Fully quantized         & -     &   80\%  \\
        Weight-only quantized   & -     &  80\% \\
        Bfloat16                & 97.93  &  81\% \\
        
  \bottomrule
\end{tabular}
\end{table}




\subsection{Impact on Performance and Energy Consumption}



Here we show the additional benefit of using AC, especially in the context of power-limited devices, such as mobile, embedded, and Edge devices.
The experiments evaluate normalized energy and delay achieved by the proposed approximate multiplier compared to a conventional baseline multiplier. Multipliers are implemented using $45~ n m$ technology via the Predictive Technology Model (PTM) using the Keysight Advanced Design System (ADS) simulation platform~\cite{PTM}.  

 \begin{table}[!htp]
 \centering
   \caption{Energy and delay Comparison.}
   \label{tab:energy}
   \begin{tabular}{ccc}
     \toprule
     \textbf{Multiplier }& \textbf{Average energy}	& \textbf{Average delay}  	 \\
     \midrule
         Exact multiplier        & 1 & 1    \\
         Ax-FPM	                 & 0.487 & 0.29\\ 
         Bfloat16                & 0.4   & 0.4   \\
   \bottomrule
 \end{tabular}
 \end{table}
 
Table \ref{tab:energy} compares the energy and delay for the approximate multiplier and the Bfloat16 multiplier normalized to a conventional multiplier. 
\begin{figure}[!htp]
\centering
\includegraphics[width=1\columnwidth]{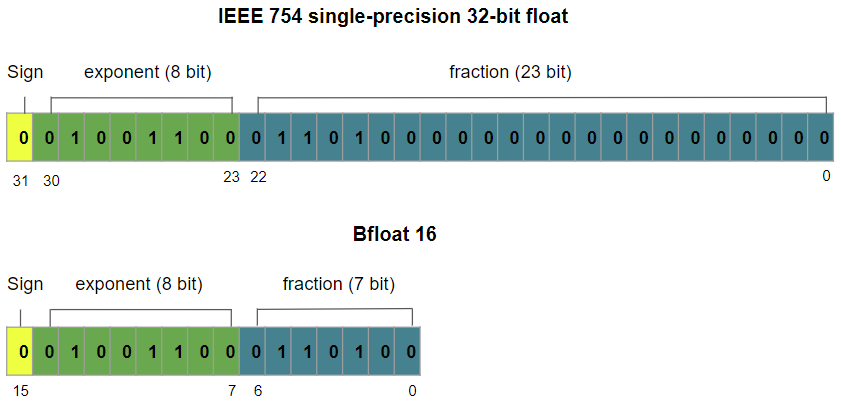}
\caption{Contrast between IEEE 754 Single-precision 32-bit floating-point format and Bfloat16 format.
}
\label{fig:bfloat}
\end{figure}

Bfloat16 data representation is shown in Figure \ref{fig:bfloat}, with 1 sign bit, a 8-bit exponent and a 7-bit fraction. The used Bfloat16 multiplier has similar architecture to that of Ax-FPM. However for the mantissa multiplier we use the conventional Booth multiplier instead of the array multiplier.
Ax-FPM achieves up to $51\%$ and $71$\% savings in energy and delay, respectively compared to the baseline multiplier. Bfloat16 implementation results in comparable savings with around $60\%$ lower power and delay. However, this comes without robustness advantages, as mentioned earlier.

 Unlike most of the state-of-the-art defense strategies that lead to power, resource, or timing overhead, DA results in saving energy and area.

\label{sec:perf_impl}

\section{Discussion}

This work tackles the problem of robustness to adversarial attacks from a new perspective: approximation in the underlying hardware. DA exploits the inherent fault tolerance of deep learning systems \cite{iccd18} to provide resilience while also obtaining the by-product gains of AC in terms of energy and resources. Our empirical study shows promising results in terms of robustness across a wide range of attack scenarios. We notice that AC-induced perturbations tend to help the classifier generalize and enhances its confidence.  We believe that this observation is possibly due to the specific AC multiplier we used where the introduced noise is input-dependent and non-uniform.
When we observe the effect on the convolution layer, we see higher absolute values when the inputs are similar to the convolution filter. This observation at the feature map propagates through the model and results in enhanced classification confidence, i.e., the difference between the $1^{st}$  and  runner-up classes. 
This aspect of confidence enhancement resembles the smoothing effect observed by some recently proposed randomization techniques~\cite{snP2019_certif}.  

 We believe that our study makes important contributions in demonstrating the general potential of approximate computing in this new dimension.  However, we believe that substantial further research remains which we hope to tackle in our future work:  (1) More work is needed to carefully understand the relationship between the patterns of induced noise and the observed robustness to adversarial attacks to guide the selection of approximation approaches; (2) We would also like to explore whether there is additional protection that results from adapting the approximation function over time; (3) We believe that DA is orthogonal to some of the other AML defenses and, deployed together, they may result in even higher protection against AML; (4) Some AML defenses unintentionally make the model more susceptible to privacy related attacks~\cite{rezaShokry}; we believe that DA does not have a similar effect and would like to study its implication on privacy-preserving in the future; and (5) We would like to explore DA in the context of other learning structures beyond CNNs.
 


DA has two important advantages compared to DQ: (1) DA results in input-dependent noise, while DQ results in a deterministic network that can be efficiently reverse engineered and undermined by adaptive white-box attacks; (2) DQ requires retraining/fine-tuning the model to avoid drastic accuracy drop, while DA does not require retraining.

Compared to DA, when proceeding to Bfloat16 quantization, we did not notice any improvement in the model robustness, which could be explained by the fact that Bfloat16 results in uniform low-amplitude noise distribution that is not sufficient to impact CNNs behavior.

While  we considered full precision floating point CNNs, we believe that DA can also apply to  quantized and sparse networks~\cite{quant,survey_121,NIPS2018_sparce} with similar impact on security. Prior work~\cite{D&T} shows that quantized networks tolerate errors, implying that DA can potentially be deployed without degrading accuracy.  

Our experimental setup is based on simulating a cross-layer implementation from gate-level up to system-level. Therefore, the experiments are  computationally and time demanding, which limited our experiments to the two datasets MNIST and CIFAR-10. This limitation was the same for techniques that require Monte Carlo Simulations such as \cite{snP2019_certif}. 
Our observations held for 5-layer-CNN (LeNet) and 8-layer-CNN (AlexNet). In future work, we are planning to solve the simulation time limitation by a real hardware implementation, which facilitate evaluating DA for larger networks.  


\label{discussion}

\section{Related Work}
Several defense mechanisms were proposed to combat adversarial attacks and can be categorized as follows:

\noindent
\textbf{Adversarial Training (AT).} 
AT is one of the most explored defenses against adversarial attacks. The main idea can be traced back to \cite{fgsm}, in which models were hardened by including adversarial examples in the training data set of the model. As a result, the trained model classifies evasive samples with higher accuracy. Various attempts to combine AT with other methods have resulted in better defense approaches such as cascade adversarial training \cite{na2017cascade}, principled training \cite{certifying}. Nonetheless, AT is not effective when the attacker uses a different attack strategy than the one used to train the model~\cite{samangouei2018defense}. Moreover, adversarial training is much more computationally intensive than training a model on the training data set only because generating evasive samples needs more computation and model fitting is more challenging (takes more epochs)~\cite{tramer2017ensemble}.


\noindent
\textbf{Input Preprocessing (IP).}
Input preprocessing depends on applying transformations to the input to remove the adversarial perturbations~\cite{das2017keeping, osadchy2017no}. Examples of transformation are denoising auto-encoders~\cite{gu2014towards}, the median, averaging, and Gaussian low-pass filters~\cite{osadchy2017no}, and JPEG compression~\cite{das2017keeping}. However, it was shown that these defenses are insecure under strong white-box attacks~\cite{chen2019towards}; if the attacker knows the specific used transformation, it can be taken into account when creating the attack. Furthermore, preprocessing requires additional computation on every input. 


\noindent
\textbf{Gradient Masking (GM).}
GM relies on applying regularization to the model to make its output less sensitive to input perturbations. Papernot et al. proposed defensive distillation~\cite{distillation_SP}, which is based on increasing the generalization of the model by distilling knowledge out of a large model to train a compact model. Nonetheless, distillation was found weak against $C\&W$ attack \cite{CW}. Nayebi and Surya~\cite{nayebi2017biologically} purposed to use saturating networks that use a loss function that promotes the activations to be in their saturating regime. \cite{ross2018improving} proposed to regularize the gradient input by penalizing variations in the model's output with respect to changes in the input during the training of differentiable models. Nonetheless, GM approaches found to make white-box attacks harder and vulnerable against black-box attacks~\cite{papernot2017practical,tramer2017ensemble}. Furthermore, they require re-training of pre-trained networks.

\noindent
\textbf{Randomization-based Defenses.} These techniques are the closest to our work \cite{snP2019_certif,smooth,stoch_prunning,liu2017,defense_certified}. Liu et al. \cite{liu2017} suggest to randomize the entire DNN and predict using an ensemble of multiple copies of the DNN. Lecuyer et al. \cite{snP2019_certif} also suggest to add random noise to the first layer of the DNN and estimate the output by a Monte Carlo simulation. These techniques offer a bounded theoretical guarantee of robustness. From a practical perspective, none of these works has been evaluated at scale or with realistic implementations.  For example, Raghunathan et al.~\cite{defense_certified} evaluate only a tiny neural network.  Other works~\cite{smooth,snP2019_certif} consider scalability but require high overhead to implement the defense (specifically, to estimate the model output which requires running a heavy Monte Carlo simulation involving a number of different runs of the CNN).
Our approach is different since not only our noise does not require overhead but comes naturally from the simpler and faster AC implementation. Moreover, while these techniques require additional training, DA is a drop-in replacement of the hardware without specific training requirements, and with no changes to the architecture nor the parameters of the CNN.  
\label{sec:rw}

\section{Conclusions}

To the best of our knowledge, this is the first work that proposes the use of hardware-supported approximation as a defense strategy against adversarial attacks for CNNs.  
We propose a CNN implementation based on Ax-FPM, an energy-efficient approximate floating-point multiplier. While AC is used in the literature to reduce the energy and delay of CNNs, we show that AC also enhances their robustness to adversarial attacks. The proposed defense is, on average, $87\%$ more robust against strong grey-box attacks and $87.5\%$ against strong black-box attacks than a conventional CNN for the case of MNIST dataset, with negligible loss in accuracy.  
The approximate CNN achieves a significant reduction in power and delay of $50\%$ and $67\%$, respectively. 



\label{conclusion}


\begin{acks}
We would like to thank the reviewers and shepherd for the feedback and suggestions that have substantially improved the paper.  This work was partially supported by NSF grants CNS-1646641, CNS-1619322 and CNS-1955650.
\end{acks}

\appendix

\section{Appendix: Design Space Exploration} \label{sec:DSE}
\textbf{Accuracy.} HEAP \cite{Heap} is the result of a design space exploration among combinations of different approximate full adders leading to minimal accuracy loss. The selection process of optimal circuit design was performed after an exhaustive accuracy evaluation of all possible configurations.
The metrics used to evaluate the accuracy of different combinations are the mean relative error distance (MRED) \cite{metrics} $ {\textstyle MRED= \frac{1}{n} \sum_{i=1}^{n} \frac{ \mid{\hat Y -Y}\mid}{Y} }$, and the normalized mean error distance (NMED): ${\textstyle NMED= \frac{1}{n} \sum_{i=1}^{n} \frac{ \mid{\hat Y -Y}\mid}{P_{max}} }$. Where $Y$ is the exact result, $ \hat Y $ is the approximate result, and $P_{max}$ is the maximum product.

\begin{table}[!htp]
\centering
  \caption{Accuracy results of the LeNet-5 CNN and different multipliers.}
  \label{tab3}
  \begin{tabular}{cccc}
    \toprule
    \textbf{Multiplier }& \textbf{CNN Accuracy}	& \textbf{MRED} & \textbf{NMED}	 \\
    \midrule
        Exact multiplier        & 97.93\%  & 0    & 0\\
        HEAP \cite{Heap}        & 97.86\%  & 0.12 & 0.03\\
        Ax-FPM	    & 97.67\%  & 0.33 & 0.08\\ 
  \bottomrule
\end{tabular}
\end{table}

Table \ref{tab3} includes the accuracy results of the exact multiplier and the two approximate multipliers.
Ax-FPM is substantially less accurate than HEAP.  However, when we evaluate CNN accuracy (using the recognition rate), we notice a negligible accuracy loss for both approximate multipliers, confirming the inherent error-resiliency of CNNs.

\begin{figure}[!htp]
\centering
\includegraphics[width=0.9\columnwidth,height=3.5cm]{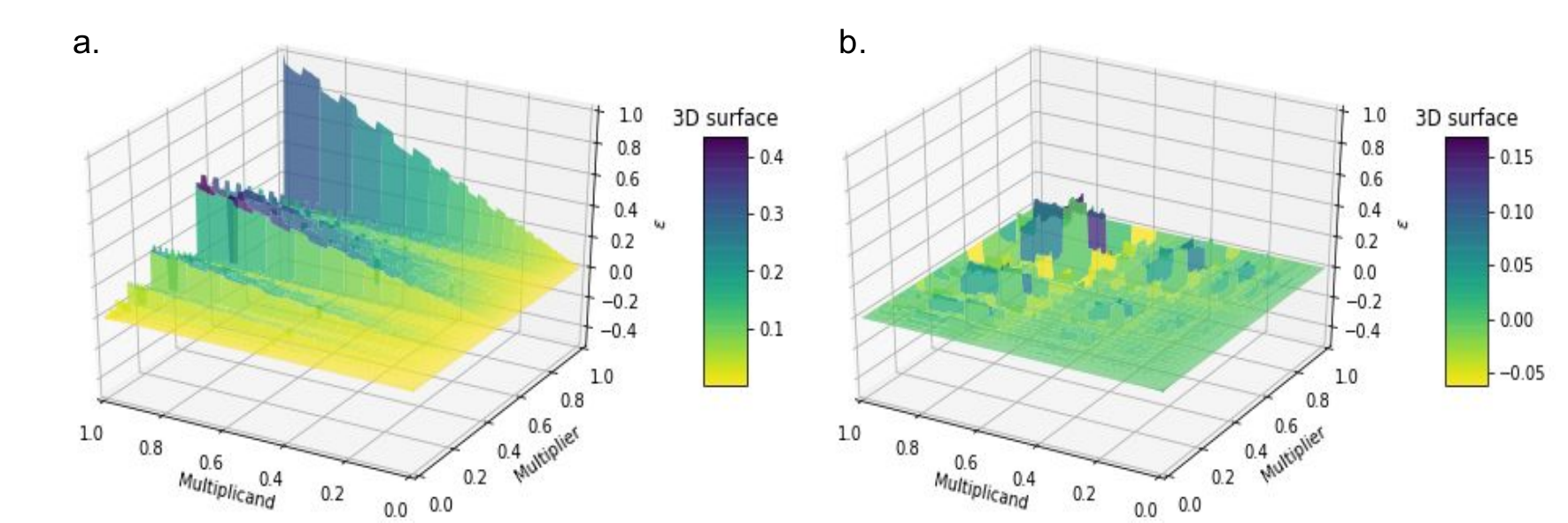}
\caption{Noise introduced by the approximate multiplier while the operands $\in [0,1]$ (a). for Ax-FPM, (b). for HEAP \cite{Heap}. 
}
\label{appx_mult_heap}
\end{figure}
Measuring the error introduced by the Ax-FPM and the HEAP for small inputs ranging between $0$ and $1$, see Figure \ref{appx_mult_heap}, we notice that, in addition to the smaller magnitude of error when using the HEAP, a different pattern was found: less data dependency and for only $34\%$ of the cases, the HEAP results were higher than the exact multiplier results. 

\begin{figure}[!htp]
\centering
\includegraphics[width=0.9\columnwidth]{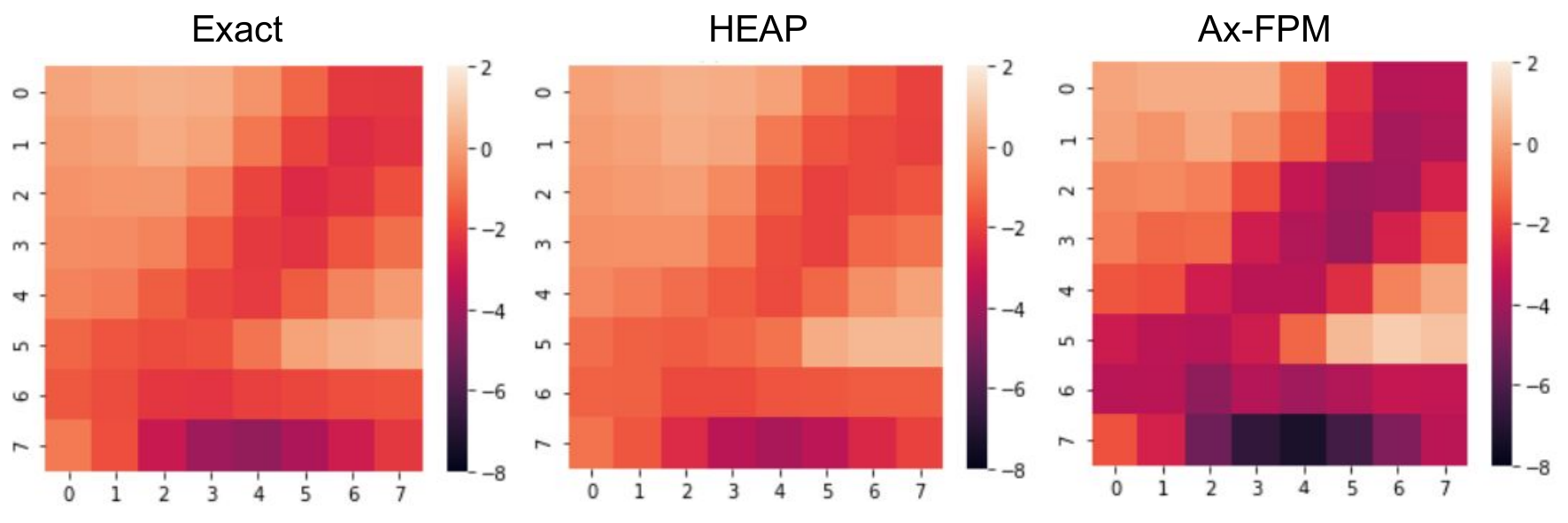}
\caption{ Heat maps of final convolution layer output when using Exact multiplier, Ax-FPM, and HEAP \cite{Heap}. }
\label{heatmaps}
\end{figure}

In order to further investigate the impact of different approximate multipliers on the convolution layer output, we produce different heat maps. As shown in Figure \ref{heatmaps}, we can see that the Ax-FPM is further highlighting the important features by increasing their scores whereas the HEAP is lowering their scores.

\noindent\textbf{Energy consumption and Delay.}
We also compare the gain in terms of energy and delay achieved by the two approximate mantissa multipliers. Multipliers are implemented using $45~ n m$ technology via the Predictive Technology Model (PTM) using the Keysight Advanced Design System (ADS) simulation platform~\cite{PTM}.   

\begin{table}[!htp]
\centering
  \caption{Energy and delay of different $24\times24$ approximate multipliers normalized to a conventional $24\times24$ multiplier.}
  \label{tab}
  \begin{tabular}{ccc}
    \toprule
    \textbf{Multiplier }& \textbf{Average energy}	& \textbf{Average delay}  	 \\
    \midrule
        Exact multiplier        & 1 & 1    \\
        HEAP                    & 0.49  & 0.46\\
        Ax-FPM	                 & 0.395 & 0.235\\ 
  \bottomrule
\end{tabular}
\end{table}

As shown in Table \ref{tab}, the mantissa multiplier in Ax-FPM achieves a considerable gain in performance and energy saving compared to the HEAP. A significant reduction in the delay was also achieved.  

\noindent\textbf{Robustness.}
We tried to assess different models transferability, the adversarial examples were generated to fool the exact model. In table \ref{grey-boxresult_with_heap}, we present the percentage of adversarial examples that successfully deceived the approximate models. While the use of the HEAP increased the robustness of the model, the Ax-FPM achieved better results.

\begin{table}[!htp]
\small
\centering
  \caption{Attacks transferability success rates for MNIST.}
  \label{grey-boxresult_with_heap}
  \begin{tabular}{cccc}
    \toprule
    \textbf{Attack} & \textbf{Exact-based}  & \textbf{HEAP-based} & \textbf{Ax-FPM-based}\\
    \midrule
        FGSM          &   100\%  & 16\% & 12\%    \\
        PGD           &   100\%  & 30\% & 28\%    \\
        JSMA          &   100\%  & 42\% & 9\%   \\
        C\&W          &   100\%  & 6\% & 1\%    \\
        DF	          &   100\%  & 18\%  & 17\%   \\ 
        LSA	          &   100\%  & 41\% & 18\%   \\ 
        BA	          &   100\%  & 22\% & 17\%   \\ 
        HSJ	          &   100\%  & 4\% & 2\%    \\         
  \bottomrule
\end{tabular}
\end{table}

All these results confirm that Ax-FPM is a better choice in terms of robustness and resource and power efficiency.

\section{Appendix: Defensive Quantization: Models Architecture}
\label{mod_archi}
We present models architectures in Table \ref{architecture}: For the fully quantized model, ConvolutionQuant, DenseQuant and reluQuant designate respectively a convolution layer with quantized weights, a dense layer with quantized weights and the relu activation function with its output quantized. For the weight quantized model architecture, the ConvolutionQuant and DenseQuant designate respectively a convolution layer with quantized weights and a dense layer with quantized weights.

\begin{table}[!htp]
\small
\centering
  \caption{Fully quantized and Weight quantized models architecture.}
  \label{architecture}
  \begin{tabular}{cc}
    \toprule
    \textbf{Fully Quantized} & \textbf{Weight Quantized}  \\
    \midrule
        ConvolutionQuant     &  ConvolutionQuant    \\
        BatchNorm            &  BatchNorm   \\
        reluQuant            &  relu    \\
        ConvolutionreluQuant &  ConvolutionreluQuant    \\
        MaxPooling           &  MaxPooling    \\
        BatchNorm            &  BatchNorm    \\
        reluQuant            &  relu  \\
        ConvolutionQuant     &  ConvolutionQuant    \\
        BatchNorm            &  BatchNorm     \\
        reluQuant            &  relu \\
        ConvolutionQuant     &  ConvolutionQuant    \\
        MaxPooling           &  MaxPooling    \\
        BatchNorm            &  BatchNorm    \\
        reluQuant            &  relu \\
        ConvolutionQuant     &  ConvolutionQuant    \\
        BatchNorm            &  BatchNorm    \\
        reluQuant            &  relu\\
        ConvolutionQuant     &  ConvolutionQuant    \\
        MaxPooling           &  MaxPooling    \\ 
        BatchNorm            &  BatchNorm    \\
        reluQuant            &  relu    \\
        DenseQuant           &  DenseQuant    \\
        BatchNorm            &  BatchNorm    \\
        reluQuant       	 &  relu   \\ 
        DenseQuant           &  DenseQuant    \\
        BatchNorm            &  BatchNorm    \\
        reluQuant	         &  relu    \\ 
        Dense                &  Dense    \\
        softmax	             &  softmax    \\ 
               
  \bottomrule
\end{tabular}
\end{table}

\balance
\bibliographystyle{ACM-Reference-Format}
\bibliography{Bib}


\begin{thebibliography}{72}


\ifx \showCODEN    \undefined \def \showCODEN     #1{\unskip}     \fi
\ifx \showDOI      \undefined \def \showDOI       #1{#1}\fi
\ifx \showISBNx    \undefined \def \showISBNx     #1{\unskip}     \fi
\ifx \showISBNxiii \undefined \def \showISBNxiii  #1{\unskip}     \fi
\ifx \showISSN     \undefined \def \showISSN      #1{\unskip}     \fi
\ifx \showLCCN     \undefined \def \showLCCN      #1{\unskip}     \fi
\ifx \shownote     \undefined \def \shownote      #1{#1}          \fi
\ifx \showarticletitle \undefined \def \showarticletitle #1{#1}   \fi
\ifx \showURL      \undefined \def \showURL       {\relax}        \fi
\providecommand\bibfield[2]{#2}
\providecommand\bibinfo[2]{#2}
\providecommand\natexlab[1]{#1}
\providecommand\showeprint[2][]{arXiv:#2}

\bibitem[\protect\citeauthoryear{??}{FP}{2008}]%
        {FP}
 \bibinfo{year}{2008}\natexlab{}.
\newblock \showarticletitle{IEEE Standard for Floating-Point Arithmetic}.
\newblock \bibinfo{journal}{\emph{IEEE Std 754-2008}} (\bibinfo{year}{2008}),
  \bibinfo{pages}{1--70}.
\newblock


\bibitem[\protect\citeauthoryear{{Akhtar} and {Mian}}{{Akhtar} and
  {Mian}}{2018}]%
        {strength}
\bibfield{author}{\bibinfo{person}{N. {Akhtar}} {and} \bibinfo{person}{A.
  {Mian}}.} \bibinfo{year}{2018}\natexlab{}.
\newblock \showarticletitle{Threat of Adversarial Attacks on Deep Learning in
  Computer Vision: A Survey}.
\newblock \bibinfo{journal}{\emph{IEEE Access}}  \bibinfo{volume}{6}
  (\bibinfo{year}{2018}), \bibinfo{pages}{14410--14430}.
\newblock


\bibitem[\protect\citeauthoryear{Al-Qizwini, Barjasteh, Al-Qassab, and
  Radha}{Al-Qizwini et~al\mbox{.}}{2017}]%
        {al2017deep}
\bibfield{author}{\bibinfo{person}{Mohammed Al-Qizwini}, \bibinfo{person}{Iman
  Barjasteh}, \bibinfo{person}{Hothaifa Al-Qassab}, {and}
  \bibinfo{person}{Hayder Radha}.} \bibinfo{year}{2017}\natexlab{}.
\newblock \showarticletitle{Deep learning algorithm for autonomous driving
  using GoogLeNet}. In \bibinfo{booktitle}{\emph{2017 IEEE Intelligent Vehicles
  Symposium (IV)}}. IEEE, \bibinfo{pages}{89--96}.
\newblock


\bibitem[\protect\citeauthoryear{Ali, Alouani, El~Cadi, Ouarnoughi, and
  Niar}{Ali et~al\mbox{.}}{2020}]%
        {quant}
\bibfield{author}{\bibinfo{person}{Karim M.~A. Ali}, \bibinfo{person}{Ihsen
  Alouani}, \bibinfo{person}{Abdessamad~Ait El~Cadi}, \bibinfo{person}{Hamza
  Ouarnoughi}, {and} \bibinfo{person}{Smail Niar}.}
  \bibinfo{year}{2020}\natexlab{}.
\newblock \showarticletitle{Cross-layer CNN Approximations for Hardware
  Implementation}. In \bibinfo{booktitle}{\emph{Applied Reconfigurable
  Computing. Architectures, Tools, and Applications}},
  \bibfield{editor}{\bibinfo{person}{Fernando Rinc{\'o}n},
  \bibinfo{person}{Jes{\'u}s Barba}, \bibinfo{person}{Hayden K.~H. So},
  \bibinfo{person}{Pedro Diniz}, {and} \bibinfo{person}{Juli{\'a}n Caba}}
  (Eds.). \bibinfo{publisher}{Springer International Publishing},
  \bibinfo{address}{Cham}, \bibinfo{pages}{151--165}.
\newblock
\showISBNx{978-3-030-44534-8}


\bibitem[\protect\citeauthoryear{{Alouani}, {Ahangari}, {Ozturk}, and
  {Niar}}{{Alouani} et~al\mbox{.}}{2018}]%
        {esl_ihsen}
\bibfield{author}{\bibinfo{person}{I. {Alouani}}, \bibinfo{person}{H.
  {Ahangari}}, \bibinfo{person}{O. {Ozturk}}, {and} \bibinfo{person}{S.
  {Niar}}.} \bibinfo{year}{2018}\natexlab{}.
\newblock \showarticletitle{A Novel Heterogeneous Approximate Multiplier for
  Low Power and High Performance}.
\newblock \bibinfo{journal}{\emph{IEEE Embedded Systems Letters}}
  \bibinfo{volume}{10}, \bibinfo{number}{2} (\bibinfo{year}{2018}),
  \bibinfo{pages}{45--48}.
\newblock


\bibitem[\protect\citeauthoryear{Betzel, Khatamifard, Suresh, Lilja, Sartori,
  and Karpuzcu}{Betzel et~al\mbox{.}}{2018}]%
        {Survey}
\bibfield{author}{\bibinfo{person}{Filipe Betzel}, \bibinfo{person}{Karen
  Khatamifard}, \bibinfo{person}{Harini Suresh}, \bibinfo{person}{David~J.
  Lilja}, \bibinfo{person}{John Sartori}, {and} \bibinfo{person}{Ulya
  Karpuzcu}.} \bibinfo{year}{2018}\natexlab{}.
\newblock \showarticletitle{Approximate Communication: Techniques for Reducing
  Communication Bottlenecks in Large-Scale Parallel Systems}.
\newblock \bibinfo{journal}{\emph{ACM Comput. Surv.}} \bibinfo{volume}{51},
  \bibinfo{number}{1}, Article \bibinfo{articleno}{1} (\bibinfo{date}{Jan.}
  \bibinfo{year}{2018}), \bibinfo{numpages}{32}~pages.
\newblock
\showISSN{0360-0300}
\urldef\tempurl%
\url{https://doi.org/10.1145/3145812}
\showDOI{\tempurl}


\bibitem[\protect\citeauthoryear{Boloor, He, Gill, Vorobeychik, and
  Zhang}{Boloor et~al\mbox{.}}{2019}]%
        {phys}
\bibfield{author}{\bibinfo{person}{Adith Boloor}, \bibinfo{person}{Xin He},
  \bibinfo{person}{Christopher Gill}, \bibinfo{person}{Yevgeniy Vorobeychik},
  {and} \bibinfo{person}{Xuan Zhang}.} \bibinfo{year}{2019}\natexlab{}.
\newblock \bibinfo{title}{Simple Physical Adversarial Examples against
  End-to-End Autonomous Driving Models}.
\newblock
\newblock
\showeprint[arxiv]{1903.05157}~[cs.RO]


\bibitem[\protect\citeauthoryear{Brendel, Rauber, and Bethge}{Brendel
  et~al\mbox{.}}{2017}]%
        {BA}
\bibfield{author}{\bibinfo{person}{Wieland Brendel}, \bibinfo{person}{Jonas
  Rauber}, {and} \bibinfo{person}{Matthias Bethge}.}
  \bibinfo{year}{2017}\natexlab{}.
\newblock \bibinfo{title}{Decision-Based Adversarial Attacks: Reliable Attacks
  Against Black-Box Machine Learning Models}.
\newblock
\newblock
\showeprint[arxiv]{1712.04248}~[stat.ML]


\bibitem[\protect\citeauthoryear{Carlini, Athalye, Papernot, Brendel, Rauber,
  Tsipras, Goodfellow, Madry, and Kurakin}{Carlini et~al\mbox{.}}{2019}]%
        {carlini_gift}
\bibfield{author}{\bibinfo{person}{Nicholas Carlini}, \bibinfo{person}{Anish
  Athalye}, \bibinfo{person}{Nicolas Papernot}, \bibinfo{person}{Wieland
  Brendel}, \bibinfo{person}{Jonas Rauber}, \bibinfo{person}{Dimitris Tsipras},
  \bibinfo{person}{Ian Goodfellow}, \bibinfo{person}{Aleksander Madry}, {and}
  \bibinfo{person}{Alexey Kurakin}.} \bibinfo{year}{2019}\natexlab{}.
\newblock \bibinfo{title}{On Evaluating Adversarial Robustness}.
\newblock
\newblock
\showeprint[arxiv]{1902.06705}~[cs.LG]


\bibitem[\protect\citeauthoryear{Carlini and Wagner}{Carlini and
  Wagner}{2016}]%
        {CW}
\bibfield{author}{\bibinfo{person}{Nicholas Carlini} {and}
  \bibinfo{person}{David~A. Wagner}.} \bibinfo{year}{2016}\natexlab{}.
\newblock \showarticletitle{Towards Evaluating the Robustness of Neural
  Networks}.
\newblock \bibinfo{journal}{\emph{CoRR}}  \bibinfo{volume}{abs/1608.04644}
  (\bibinfo{year}{2016}).
\newblock
\showeprint[arxiv]{1608.04644}
\urldef\tempurl%
\url{http://arxiv.org/abs/1608.04644}
\showURL{%
\tempurl}


\bibitem[\protect\citeauthoryear{Chen and Jordan}{Chen and Jordan}{2019}]%
        {HSJ}
\bibfield{author}{\bibinfo{person}{Jianbo Chen} {and}
  \bibinfo{person}{Michael~I. Jordan}.} \bibinfo{year}{2019}\natexlab{}.
\newblock \showarticletitle{Boundary Attack++: Query-Efficient Decision-Based
  Adversarial Attack}.
\newblock \bibinfo{journal}{\emph{CoRR}}  \bibinfo{volume}{abs/1904.02144}
  (\bibinfo{year}{2019}).
\newblock
\showeprint[arxiv]{1904.02144}
\urldef\tempurl%
\url{http://arxiv.org/abs/1904.02144}
\showURL{%
\tempurl}


\bibitem[\protect\citeauthoryear{Chen, Wu, Rastogi, Liang, and Jha}{Chen
  et~al\mbox{.}}{2019}]%
        {chen2019towards}
\bibfield{author}{\bibinfo{person}{Jiefeng Chen}, \bibinfo{person}{Xi Wu},
  \bibinfo{person}{Vaibhav Rastogi}, \bibinfo{person}{Yingyu Liang}, {and}
  \bibinfo{person}{Somesh Jha}.} \bibinfo{year}{2019}\natexlab{}.
\newblock \showarticletitle{Towards understanding limitations of pixel
  discretization against adversarial attacks}. In
  \bibinfo{booktitle}{\emph{2019 IEEE European Symposium on Security and
  Privacy (EuroS\&P)}}. IEEE, \bibinfo{pages}{480--495}.
\newblock


\bibitem[\protect\citeauthoryear{Ciresan, Meier, Masci, and
  Schmidhuber}{Ciresan et~al\mbox{.}}{2012}]%
        {signs}
\bibfield{author}{\bibinfo{person}{Dan~C. Ciresan}, \bibinfo{person}{Ueli
  Meier}, \bibinfo{person}{Jonathan Masci}, {and} \bibinfo{person}{J{\"u}rgen
  Schmidhuber}.} \bibinfo{year}{2012}\natexlab{}.
\newblock \showarticletitle{Multi-column deep neural network for traffic sign
  classification}.
\newblock \bibinfo{journal}{\emph{Neural networks : the official journal of the
  International Neural Network Society}}  \bibinfo{volume}{32}
  (\bibinfo{year}{2012}), \bibinfo{pages}{333--8}.
\newblock


\bibitem[\protect\citeauthoryear{Cohen, Rosenfeld, and Kolter}{Cohen
  et~al\mbox{.}}{2019}]%
        {smooth}
\bibfield{author}{\bibinfo{person}{Jeremy Cohen}, \bibinfo{person}{Elan
  Rosenfeld}, {and} \bibinfo{person}{Zico Kolter}.}
  \bibinfo{year}{2019}\natexlab{}.
\newblock \showarticletitle{Certified Adversarial Robustness via Randomized
  Smoothing}. In \bibinfo{booktitle}{\emph{Proceedings of the 36th
  International Conference on Machine Learning}}
  \emph{(\bibinfo{series}{Proceedings of Machine Learning Research},
  Vol.~\bibinfo{volume}{97})}, \bibfield{editor}{\bibinfo{person}{Kamalika
  Chaudhuri} {and} \bibinfo{person}{Ruslan Salakhutdinov}} (Eds.).
  \bibinfo{publisher}{PMLR}, \bibinfo{address}{Long Beach, California, USA},
  \bibinfo{pages}{1310--1320}.
\newblock
\urldef\tempurl%
\url{http://proceedings.mlr.press/v97/cohen19c.html}
\showURL{%
\tempurl}


\bibitem[\protect\citeauthoryear{Das, Shanbhogue, Chen, Hohman, Chen, Kounavis,
  and Chau}{Das et~al\mbox{.}}{2017}]%
        {das2017keeping}
\bibfield{author}{\bibinfo{person}{Nilaksh Das}, \bibinfo{person}{Madhuri
  Shanbhogue}, \bibinfo{person}{Shang-Tse Chen}, \bibinfo{person}{Fred Hohman},
  \bibinfo{person}{Li Chen}, \bibinfo{person}{Michael~E Kounavis}, {and}
  \bibinfo{person}{Duen~Horng Chau}.} \bibinfo{year}{2017}\natexlab{}.
\newblock \showarticletitle{Keeping the bad guys out: Protecting and
  vaccinating deep learning with jpeg compression}.
\newblock \bibinfo{journal}{\emph{arXiv preprint arXiv:1705.02900}}
  (\bibinfo{year}{2017}).
\newblock


\bibitem[\protect\citeauthoryear{Deng and Liu}{Deng and Liu}{2018}]%
        {deng2018deep}
\bibfield{author}{\bibinfo{person}{Li Deng} {and} \bibinfo{person}{Yang Liu}.}
  \bibinfo{year}{2018}\natexlab{}.
\newblock \bibinfo{booktitle}{\emph{Deep learning in natural language
  processing}}.
\newblock \bibinfo{publisher}{Springer}.
\newblock


\bibitem[\protect\citeauthoryear{Deng}{Deng}{2019}]%
        {reviewDL}
\bibfield{author}{\bibinfo{person}{Yunbin Deng}.}
  \bibinfo{year}{2019}\natexlab{}.
\newblock \showarticletitle{Deep Learning on Mobile Devices - {A} Review}.
\newblock \bibinfo{journal}{\emph{CoRR}}  \bibinfo{volume}{abs/1904.09274}
  (\bibinfo{year}{2019}).
\newblock
\showeprint[arxiv]{1904.09274}
\urldef\tempurl%
\url{http://arxiv.org/abs/1904.09274}
\showURL{%
\tempurl}


\bibitem[\protect\citeauthoryear{Dhillon, Azizzadenesheli, Lipton, Bernstein,
  Kossaifi, Khanna, and Anandkumar}{Dhillon et~al\mbox{.}}{2018}]%
        {stoch_prunning}
\bibfield{author}{\bibinfo{person}{Guneet~S. Dhillon}, \bibinfo{person}{Kamyar
  Azizzadenesheli}, \bibinfo{person}{Zachary~C. Lipton},
  \bibinfo{person}{Jeremy Bernstein}, \bibinfo{person}{Jean Kossaifi},
  \bibinfo{person}{Aran Khanna}, {and} \bibinfo{person}{Anima Anandkumar}.}
  \bibinfo{year}{2018}\natexlab{}.
\newblock \showarticletitle{Stochastic Activation Pruning for Robust
  Adversarial Defense}.
\newblock \bibinfo{journal}{\emph{CoRR}}  \bibinfo{volume}{abs/1803.01442}
  (\bibinfo{year}{2018}).
\newblock
\showeprint[arxiv]{1803.01442}
\urldef\tempurl%
\url{http://arxiv.org/abs/1803.01442}
\showURL{%
\tempurl}


\bibitem[\protect\citeauthoryear{Evtimov, Eykholt, Fernandes, Kohno, Li,
  Prakash, Rahmati, and Song}{Evtimov et~al\mbox{.}}{2017}]%
        {phy10}
\bibfield{author}{\bibinfo{person}{Ivan Evtimov}, \bibinfo{person}{Kevin
  Eykholt}, \bibinfo{person}{Earlence Fernandes}, \bibinfo{person}{Tadayoshi
  Kohno}, \bibinfo{person}{Bo Li}, \bibinfo{person}{Atul Prakash},
  \bibinfo{person}{Amir Rahmati}, {and} \bibinfo{person}{Dawn Song}.}
  \bibinfo{year}{2017}\natexlab{}.
\newblock \showarticletitle{Robust Physical-World Attacks on Machine Learning
  Models}.
\newblock \bibinfo{journal}{\emph{CoRR}}  \bibinfo{volume}{abs/1707.08945}
  (\bibinfo{year}{2017}).
\newblock
\showeprint[arxiv]{1707.08945}
\urldef\tempurl%
\url{http://arxiv.org/abs/1707.08945}
\showURL{%
\tempurl}


\bibitem[\protect\citeauthoryear{Goodfellow, Shlens, and Szegedy}{Goodfellow
  et~al\mbox{.}}{2014}]%
        {fgsm}
\bibfield{author}{\bibinfo{person}{Ian~J. Goodfellow},
  \bibinfo{person}{Jonathon Shlens}, {and} \bibinfo{person}{Christian
  Szegedy}.} \bibinfo{year}{2014}\natexlab{}.
\newblock \bibinfo{title}{Explaining and Harnessing Adversarial Examples}.
\newblock
\newblock
\showeprint[arxiv]{1412.6572}~[stat.ML]


\bibitem[\protect\citeauthoryear{Gu and Rigazio}{Gu and Rigazio}{2014}]%
        {gu2014towards}
\bibfield{author}{\bibinfo{person}{Shixiang Gu} {and} \bibinfo{person}{Luca
  Rigazio}.} \bibinfo{year}{2014}\natexlab{}.
\newblock \showarticletitle{Towards deep neural network architectures robust to
  adversarial examples}.
\newblock \bibinfo{journal}{\emph{arXiv preprint arXiv:1412.5068}}
  (\bibinfo{year}{2014}).
\newblock


\bibitem[\protect\citeauthoryear{Guesmi, Alouani, Baklouti, Frikha, Abid, and
  Rivenq}{Guesmi et~al\mbox{.}}{2019}]%
        {Heap}
\bibfield{author}{\bibinfo{person}{Amira Guesmi}, \bibinfo{person}{Ihsen
  Alouani}, \bibinfo{person}{Mouna Baklouti}, \bibinfo{person}{Tarek Frikha},
  \bibinfo{person}{Mohamed Abid}, {and} \bibinfo{person}{Atika Rivenq}.}
  \bibinfo{year}{2019}\natexlab{}.
\newblock \showarticletitle{HEAP: A Heterogeneous Approximate Floating-Point
  Multiplier for Error Tolerant Applications}. In
  \bibinfo{booktitle}{\emph{Proceedings of the 30th International Workshop on
  Rapid System Prototyping (RSP'19)}} (New York, NY, USA)
  \emph{(\bibinfo{series}{RSP '19})}. \bibinfo{publisher}{ACM},
  \bibinfo{address}{New York, NY, USA}, \bibinfo{pages}{36--42}.
\newblock
\showISBNx{978-1-4503-6847-6}
\urldef\tempurl%
\url{https://doi.org/10.1145/3339985.3358495}
\showDOI{\tempurl}


\bibitem[\protect\citeauthoryear{{Gupta}, {Mohapatra}, {Raghunathan}, and
  {Roy}}{{Gupta} et~al\mbox{.}}{2013}]%
        {AMA5}
\bibfield{author}{\bibinfo{person}{V. {Gupta}}, \bibinfo{person}{D.
  {Mohapatra}}, \bibinfo{person}{A. {Raghunathan}}, {and} \bibinfo{person}{K.
  {Roy}}.} \bibinfo{year}{2013}\natexlab{}.
\newblock \showarticletitle{Low-Power Digital Signal Processing Using
  Approximate Adders}.
\newblock \bibinfo{journal}{\emph{IEEE Transactions on Computer-Aided Design of
  Integrated Circuits and Systems}} \bibinfo{volume}{32}, \bibinfo{number}{1}
  (\bibinfo{date}{Jan} \bibinfo{year}{2013}), \bibinfo{pages}{124--137}.
\newblock
\urldef\tempurl%
\url{https://doi.org/10.1109/TCAD.2012.2217962}
\showDOI{\tempurl}


\bibitem[\protect\citeauthoryear{He, Zhang, Ren, and Sun}{He
  et~al\mbox{.}}{2015}]%
        {He}
\bibfield{author}{\bibinfo{person}{Kaiming He}, \bibinfo{person}{Xiangyu
  Zhang}, \bibinfo{person}{Shaoqing Ren}, {and} \bibinfo{person}{Jian Sun}.}
  \bibinfo{year}{2015}\natexlab{}.
\newblock \showarticletitle{Deep Residual Learning for Image Recognition}.
\newblock \bibinfo{journal}{\emph{CoRR}}  \bibinfo{volume}{abs/1512.03385}
  (\bibinfo{year}{2015}).
\newblock
\showeprint[arxiv]{1512.03385}
\urldef\tempurl%
\url{http://arxiv.org/abs/1512.03385}
\showURL{%
\tempurl}


\bibitem[\protect\citeauthoryear{Hendrycks and Gimpel}{Hendrycks and
  Gimpel}{2016}]%
        {early}
\bibfield{author}{\bibinfo{person}{Dan Hendrycks} {and} \bibinfo{person}{Kevin
  Gimpel}.} \bibinfo{year}{2016}\natexlab{}.
\newblock \bibinfo{title}{Early Methods for Detecting Adversarial Images}.
\newblock
\newblock
\showeprint[arxiv]{1608.00530}~[cs.LG]


\bibitem[\protect\citeauthoryear{{Hrbacek}, {Mrazek}, and {Vasicek}}{{Hrbacek}
  et~al\mbox{.}}{2016}]%
        {array}
\bibfield{author}{\bibinfo{person}{R. {Hrbacek}}, \bibinfo{person}{V.
  {Mrazek}}, {and} \bibinfo{person}{Z. {Vasicek}}.}
  \bibinfo{year}{2016}\natexlab{}.
\newblock \showarticletitle{Automatic design of approximate circuits by means
  of multi-objective evolutionary algorithms}. In
  \bibinfo{booktitle}{\emph{2016 International Conference on Design and
  Technology of Integrated Systems in Nanoscale Era (DTIS)}}.
  \bibinfo{pages}{1--6}.
\newblock
\urldef\tempurl%
\url{https://doi.org/10.1109/DTIS.2016.7483885}
\showDOI{\tempurl}


\bibitem[\protect\citeauthoryear{Integration and Group}{Integration and
  Group}{2012}]%
        {PTM}
\bibfield{author}{\bibinfo{person}{Nanoscale Integration} {and}
  \bibinfo{person}{Modeling~(NIMO) Group}.} \bibinfo{year}{2012}\natexlab{}.
\newblock \bibinfo{booktitle}{\emph{Predictive technology model (PTM)
  website}}.
\newblock
\urldef\tempurl%
\url{http://ptm.asu.edu}
\showURL{%
Retrieved April 8, 2019 from \tempurl}


\bibitem[\protect\citeauthoryear{Kalamkar, Mudigere, Mellempudi, Das, Banerjee,
  Avancha, Vooturi, Jammalamadaka, Huang, Yuen, Yang, Park, Heinecke,
  Georganas, Srinivasan, Kundu, Smelyanskiy, Kaul, and Dubey}{Kalamkar
  et~al\mbox{.}}{2019}]%
        {kalamkar2019study}
\bibfield{author}{\bibinfo{person}{Dhiraj Kalamkar}, \bibinfo{person}{Dheevatsa
  Mudigere}, \bibinfo{person}{Naveen Mellempudi}, \bibinfo{person}{Dipankar
  Das}, \bibinfo{person}{Kunal Banerjee}, \bibinfo{person}{Sasikanth Avancha},
  \bibinfo{person}{Dharma~Teja Vooturi}, \bibinfo{person}{Nataraj
  Jammalamadaka}, \bibinfo{person}{Jianyu Huang}, \bibinfo{person}{Hector
  Yuen}, \bibinfo{person}{Jiyan Yang}, \bibinfo{person}{Jongsoo Park},
  \bibinfo{person}{Alexander Heinecke}, \bibinfo{person}{Evangelos Georganas},
  \bibinfo{person}{Sudarshan Srinivasan}, \bibinfo{person}{Abhisek Kundu},
  \bibinfo{person}{Misha Smelyanskiy}, \bibinfo{person}{Bharat Kaul}, {and}
  \bibinfo{person}{Pradeep Dubey}.} \bibinfo{year}{2019}\natexlab{}.
\newblock \bibinfo{title}{A Study of BFLOAT16 for Deep Learning Training}.
\newblock
\newblock
\showeprint[arxiv]{1905.12322}~[cs.LG]


\bibitem[\protect\citeauthoryear{Krizhevsky}{Krizhevsky}{2009}]%
        {CIFAR}
\bibfield{author}{\bibinfo{person}{Alex Krizhevsky}.}
  \bibinfo{year}{2009}\natexlab{}.
\newblock \bibinfo{booktitle}{\emph{Learning multiple layers of features from
  tiny images}}.
\newblock \bibinfo{type}{{T}echnical {R}eport}.
\newblock


\bibitem[\protect\citeauthoryear{Krizhevsky, Sutskever, and Hinton}{Krizhevsky
  et~al\mbox{.}}{2012}]%
        {Alexnet}
\bibfield{author}{\bibinfo{person}{Alex Krizhevsky}, \bibinfo{person}{Ilya
  Sutskever}, {and} \bibinfo{person}{Geoffrey~E. Hinton}.}
  \bibinfo{year}{2012}\natexlab{}.
\newblock \showarticletitle{ImageNet Classification with Deep Convolutional
  Neural Networks}. In \bibinfo{booktitle}{\emph{Proceedings of the 25th
  International Conference on Neural Information Processing Systems - Volume
  1}} (Lake Tahoe, Nevada) \emph{(\bibinfo{series}{NIPS’12})}.
  \bibinfo{publisher}{Curran Associates Inc.}, \bibinfo{address}{Red Hook, NY,
  USA}, \bibinfo{pages}{1097–1105}.
\newblock


\bibitem[\protect\citeauthoryear{Kurakin, Goodfellow, and Bengio}{Kurakin
  et~al\mbox{.}}{2016}]%
        {phy9}
\bibfield{author}{\bibinfo{person}{Alexey Kurakin}, \bibinfo{person}{Ian~J.
  Goodfellow}, {and} \bibinfo{person}{Samy Bengio}.}
  \bibinfo{year}{2016}\natexlab{}.
\newblock \showarticletitle{Adversarial examples in the physical world}.
\newblock \bibinfo{journal}{\emph{CoRR}}  \bibinfo{volume}{abs/1607.02533}
  (\bibinfo{year}{2016}).
\newblock
\showeprint[arxiv]{1607.02533}
\urldef\tempurl%
\url{http://arxiv.org/abs/1607.02533}
\showURL{%
\tempurl}


\bibitem[\protect\citeauthoryear{{Lecun}, {Bottou}, {Bengio}, and
  {Haffner}}{{Lecun} et~al\mbox{.}}{1998}]%
        {lenet5}
\bibfield{author}{\bibinfo{person}{Y. {Lecun}}, \bibinfo{person}{L. {Bottou}},
  \bibinfo{person}{Y. {Bengio}}, {and} \bibinfo{person}{P. {Haffner}}.}
  \bibinfo{year}{1998}\natexlab{}.
\newblock \showarticletitle{Gradient-based learning applied to document
  recognition}.
\newblock \bibinfo{journal}{\emph{Proc. IEEE}} \bibinfo{volume}{86},
  \bibinfo{number}{11} (\bibinfo{date}{Nov} \bibinfo{year}{1998}),
  \bibinfo{pages}{2278--2324}.
\newblock
\urldef\tempurl%
\url{https://doi.org/10.1109/5.726791}
\showDOI{\tempurl}


\bibitem[\protect\citeauthoryear{LeCun and Cortes}{LeCun and Cortes}{2010}]%
        {mnist}
\bibfield{author}{\bibinfo{person}{Yann LeCun} {and} \bibinfo{person}{Corinna
  Cortes}.} \bibinfo{year}{2010}\natexlab{}.
\newblock \showarticletitle{{MNIST} handwritten digit database}.
\newblock \bibinfo{howpublished}{http://yann.lecun.com/exdb/mnist/}.
\newblock  (\bibinfo{year}{2010}).
\newblock
\urldef\tempurl%
\url{http://yann.lecun.com/exdb/mnist/}
\showURL{%
\tempurl}


\bibitem[\protect\citeauthoryear{{Lecuyer}, {Atlidakis}, {Geambasu}, {Hsu}, and
  {Jana}}{{Lecuyer} et~al\mbox{.}}{2019}]%
        {snP2019_certif}
\bibfield{author}{\bibinfo{person}{M. {Lecuyer}}, \bibinfo{person}{V.
  {Atlidakis}}, \bibinfo{person}{R. {Geambasu}}, \bibinfo{person}{D. {Hsu}},
  {and} \bibinfo{person}{S. {Jana}}.} \bibinfo{year}{2019}\natexlab{}.
\newblock \showarticletitle{Certified Robustness to Adversarial Examples with
  Differential Privacy}. In \bibinfo{booktitle}{\emph{2019 IEEE Symposium on
  Security and Privacy (SP)}}. \bibinfo{pages}{656--672}.
\newblock


\bibitem[\protect\citeauthoryear{{Liang}, {Han}, and {Lombardi}}{{Liang}
  et~al\mbox{.}}{2013}]%
        {metrics}
\bibfield{author}{\bibinfo{person}{J. {Liang}}, \bibinfo{person}{J. {Han}},
  {and} \bibinfo{person}{F. {Lombardi}}.} \bibinfo{year}{2013}\natexlab{}.
\newblock \showarticletitle{New Metrics for the Reliability of Approximate and
  Probabilistic Adders}.
\newblock \bibinfo{journal}{\emph{IEEE Trans. Comput.}} \bibinfo{volume}{62},
  \bibinfo{number}{9} (\bibinfo{date}{Sep.} \bibinfo{year}{2013}),
  \bibinfo{pages}{1760--1771}.
\newblock
\showISSN{0018-9340}
\urldef\tempurl%
\url{https://doi.org/10.1109/TC.2012.146}
\showDOI{\tempurl}


\bibitem[\protect\citeauthoryear{Lin, Gan, and Han}{Lin et~al\mbox{.}}{2019}]%
        {DQ}
\bibfield{author}{\bibinfo{person}{Ji Lin}, \bibinfo{person}{Chuang Gan}, {and}
  \bibinfo{person}{Song Han}.} \bibinfo{year}{2019}\natexlab{}.
\newblock \bibinfo{title}{Defensive Quantization: When Efficiency Meets
  Robustness}.
\newblock
\newblock
\showeprint[arxiv]{1904.08444}~[cs.LG]


\bibitem[\protect\citeauthoryear{Liu, Cheng, Zhang, and Hsieh}{Liu
  et~al\mbox{.}}{2017}]%
        {liu2017}
\bibfield{author}{\bibinfo{person}{Xuanqing Liu}, \bibinfo{person}{Minhao
  Cheng}, \bibinfo{person}{Huan Zhang}, {and} \bibinfo{person}{Cho-Jui Hsieh}.}
  \bibinfo{year}{2017}\natexlab{}.
\newblock \bibinfo{title}{Towards Robust Neural Networks via Random
  Self-ensemble}.
\newblock
\newblock
\showeprint[arxiv]{1712.00673}~[cs.LG]


\bibitem[\protect\citeauthoryear{Lu, Sibai, Fabry, and Forsyth}{Lu
  et~al\mbox{.}}{2017}]%
        {noneed}
\bibfield{author}{\bibinfo{person}{Jiajun Lu}, \bibinfo{person}{Hussein Sibai},
  \bibinfo{person}{Evan Fabry}, {and} \bibinfo{person}{David Forsyth}.}
  \bibinfo{year}{2017}\natexlab{}.
\newblock \bibinfo{title}{NO Need to Worry about Adversarial Examples in Object
  Detection in Autonomous Vehicles}.
\newblock
\newblock
\showeprint[arxiv]{1707.03501}~[cs.CV]


\bibitem[\protect\citeauthoryear{Ma, Suda, Cao, Seo, and Vrudhula}{Ma
  et~al\mbox{.}}{2016}]%
        {survey_121}
\bibfield{author}{\bibinfo{person}{Yufei Ma}, \bibinfo{person}{N. Suda},
  \bibinfo{person}{Yu Cao}, \bibinfo{person}{J. Seo}, {and} \bibinfo{person}{S.
  Vrudhula}.} \bibinfo{year}{2016}\natexlab{}.
\newblock \showarticletitle{Scalable and modularized RTL compilation of
  Convolutional Neural Networks onto FPGA}. In \bibinfo{booktitle}{\emph{2016
  26th International Conference on Field Programmable Logic and Applications
  (FPL)}}. \bibinfo{pages}{1--8}.
\newblock
\showISSN{1946-1488}


\bibitem[\protect\citeauthoryear{Madry, Makelov, Schmidt, Tsipras, and
  Vladu}{Madry et~al\mbox{.}}{2017a}]%
        {DLres}
\bibfield{author}{\bibinfo{person}{Aleksander Madry},
  \bibinfo{person}{Aleksandar Makelov}, \bibinfo{person}{Ludwig Schmidt},
  \bibinfo{person}{Dimitris Tsipras}, {and} \bibinfo{person}{Adrian Vladu}.}
  \bibinfo{year}{2017}\natexlab{a}.
\newblock \bibinfo{title}{Towards Deep Learning Models Resistant to Adversarial
  Attacks}.
\newblock
\newblock
\showeprint[arxiv]{1706.06083}~[stat.ML]


\bibitem[\protect\citeauthoryear{Madry, Makelov, Schmidt, Tsipras, and
  Vladu}{Madry et~al\mbox{.}}{2017b}]%
        {pgd}
\bibfield{author}{\bibinfo{person}{Aleksander Madry},
  \bibinfo{person}{Aleksandar Makelov}, \bibinfo{person}{Ludwig Schmidt},
  \bibinfo{person}{Dimitris Tsipras}, {and} \bibinfo{person}{Adrian Vladu}.}
  \bibinfo{year}{2017}\natexlab{b}.
\newblock \bibinfo{title}{Towards Deep Learning Models Resistant to Adversarial
  Attacks}.
\newblock
\newblock
\showeprint[arxiv]{1706.06083}~[stat.ML]


\bibitem[\protect\citeauthoryear{Miah, Yousuf, Mia, and Miya}{Miah
  et~al\mbox{.}}{2015}]%
        {cheque}
\bibfield{author}{\bibinfo{person}{Mohammad Badrul~Alam Miah},
  \bibinfo{person}{Mohammad~Abu Yousuf}, \bibinfo{person}{Md.~Sohag Mia}, {and}
  \bibinfo{person}{Md.~Parag Miya}.} \bibinfo{year}{2015}\natexlab{}.
\newblock \showarticletitle{Article: Handwritten Courtesy Amount and Signature
  Recognition on Bank Cheque using Neural Network}.
\newblock \bibinfo{journal}{\emph{International Journal of Computer
  Applications}} \bibinfo{volume}{118}, \bibinfo{number}{5}
  (\bibinfo{date}{May} \bibinfo{year}{2015}), \bibinfo{pages}{21--26}.
\newblock
\newblock
\shownote{Full text available.}


\bibitem[\protect\citeauthoryear{Miotto, Wang, Wang, Jiang, and Dudley}{Miotto
  et~al\mbox{.}}{2018}]%
        {miotto2018deep}
\bibfield{author}{\bibinfo{person}{Riccardo Miotto}, \bibinfo{person}{Fei
  Wang}, \bibinfo{person}{Shuang Wang}, \bibinfo{person}{Xiaoqian Jiang}, {and}
  \bibinfo{person}{Joel~T Dudley}.} \bibinfo{year}{2018}\natexlab{}.
\newblock \showarticletitle{Deep learning for healthcare: review, opportunities
  and challenges}.
\newblock \bibinfo{journal}{\emph{Briefings in bioinformatics}}
  \bibinfo{volume}{19}, \bibinfo{number}{6} (\bibinfo{year}{2018}),
  \bibinfo{pages}{1236--1246}.
\newblock


\bibitem[\protect\citeauthoryear{{Moore}}{{Moore}}{2019}]%
        {tsmc_moore}
\bibfield{author}{\bibinfo{person}{S.~K. {Moore}}.}
  \bibinfo{year}{2019}\natexlab{}.
\newblock \showarticletitle{Another step toward the end of Moore's law: Samsung
  and TSMC move to 5-nanometer manufacturing - [News]}.
\newblock \bibinfo{journal}{\emph{IEEE Spectrum}} \bibinfo{volume}{56},
  \bibinfo{number}{6} (\bibinfo{date}{June} \bibinfo{year}{2019}),
  \bibinfo{pages}{9--10}.
\newblock
\showISSN{1939-9340}
\urldef\tempurl%
\url{https://doi.org/10.1109/MSPEC.2019.8727133}
\showDOI{\tempurl}


\bibitem[\protect\citeauthoryear{Moosavi-Dezfooli, Fawzi, and
  Frossard}{Moosavi-Dezfooli et~al\mbox{.}}{2015}]%
        {deepfool}
\bibfield{author}{\bibinfo{person}{Seyed-Mohsen Moosavi-Dezfooli},
  \bibinfo{person}{Alhussein Fawzi}, {and} \bibinfo{person}{Pascal Frossard}.}
  \bibinfo{year}{2015}\natexlab{}.
\newblock \bibinfo{title}{DeepFool: a simple and accurate method to fool deep
  neural networks}.
\newblock
\newblock
\showeprint[arxiv]{1511.04599}~[cs.LG]


\bibitem[\protect\citeauthoryear{Na, Ko, and Mukhopadhyay}{Na
  et~al\mbox{.}}{2017}]%
        {na2017cascade}
\bibfield{author}{\bibinfo{person}{Taesik Na}, \bibinfo{person}{Jong~Hwan Ko},
  {and} \bibinfo{person}{Saibal Mukhopadhyay}.}
  \bibinfo{year}{2017}\natexlab{}.
\newblock \bibinfo{title}{Cascade Adversarial Machine Learning Regularized with
  a Unified Embedding}.
\newblock
\newblock
\showeprint[arxiv]{1708.02582}~[stat.ML]


\bibitem[\protect\citeauthoryear{Narodytska and Kasiviswanathan}{Narodytska and
  Kasiviswanathan}{2016}]%
        {localsearch}
\bibfield{author}{\bibinfo{person}{Nina Narodytska} {and}
  \bibinfo{person}{Shiva~Prasad Kasiviswanathan}.}
  \bibinfo{year}{2016}\natexlab{}.
\newblock \showarticletitle{Simple Black-Box Adversarial Perturbations for Deep
  Networks}.
\newblock \bibinfo{journal}{\emph{CoRR}}  \bibinfo{volume}{abs/1612.06299}
  (\bibinfo{year}{2016}).
\newblock
\showeprint[arxiv]{1612.06299}
\urldef\tempurl%
\url{http://arxiv.org/abs/1612.06299}
\showURL{%
\tempurl}


\bibitem[\protect\citeauthoryear{Nayebi and Ganguli}{Nayebi and
  Ganguli}{2017}]%
        {nayebi2017biologically}
\bibfield{author}{\bibinfo{person}{Aran Nayebi} {and} \bibinfo{person}{Surya
  Ganguli}.} \bibinfo{year}{2017}\natexlab{}.
\newblock \bibinfo{title}{Biologically inspired protection of deep networks
  from adversarial attacks}.
\newblock
\newblock
\showeprint[arxiv]{1703.09202}~[stat.ML]


\bibitem[\protect\citeauthoryear{{Neggaz}, {Alouani}, {Lorenzo}, and
  {Niar}}{{Neggaz} et~al\mbox{.}}{2018}]%
        {iccd18}
\bibfield{author}{\bibinfo{person}{M.~A. {Neggaz}}, \bibinfo{person}{I.
  {Alouani}}, \bibinfo{person}{P.~R. {Lorenzo}}, {and} \bibinfo{person}{S.
  {Niar}}.} \bibinfo{year}{2018}\natexlab{}.
\newblock \showarticletitle{A Reliability Study on CNNs for Critical Embedded
  Systems}. In \bibinfo{booktitle}{\emph{2018 IEEE 36th International
  Conference on Computer Design (ICCD)}}. \bibinfo{pages}{476--479}.
\newblock


\bibitem[\protect\citeauthoryear{{Neggaz}, {Alouani}, {Niar}, and
  {Kurdahi}}{{Neggaz} et~al\mbox{.}}{2019}]%
        {D&T}
\bibfield{author}{\bibinfo{person}{M.~A. {Neggaz}}, \bibinfo{person}{I.
  {Alouani}}, \bibinfo{person}{S. {Niar}}, {and} \bibinfo{person}{F.
  {Kurdahi}}.} \bibinfo{year}{2019}\natexlab{}.
\newblock \showarticletitle{Are CNNs Reliable Enough for Critical Applications?
  An Exploratory Study}.
\newblock \bibinfo{journal}{\emph{IEEE Design Test}} (\bibinfo{year}{2019}),
  \bibinfo{pages}{1--1}.
\newblock
\showISSN{2168-2364}
\urldef\tempurl%
\url{https://doi.org/10.1109/MDAT.2019.2952336}
\showDOI{\tempurl}


\bibitem[\protect\citeauthoryear{Osadchy, Hernandez-Castro, Gibson, Dunkelman,
  and P{\'e}rez-Cabo}{Osadchy et~al\mbox{.}}{2017}]%
        {osadchy2017no}
\bibfield{author}{\bibinfo{person}{Margarita Osadchy}, \bibinfo{person}{Julio
  Hernandez-Castro}, \bibinfo{person}{Stuart Gibson}, \bibinfo{person}{Orr
  Dunkelman}, {and} \bibinfo{person}{Daniel P{\'e}rez-Cabo}.}
  \bibinfo{year}{2017}\natexlab{}.
\newblock \showarticletitle{No bot expects the DeepCAPTCHA! Introducing
  immutable adversarial examples, with applications to CAPTCHA generation}.
\newblock \bibinfo{journal}{\emph{IEEE Transactions on Information Forensics
  and Security}} \bibinfo{volume}{12}, \bibinfo{number}{11}
  (\bibinfo{year}{2017}), \bibinfo{pages}{2640--2653}.
\newblock


\bibitem[\protect\citeauthoryear{Papernot, McDaniel, Goodfellow, Jha, Celik,
  and Swami}{Papernot et~al\mbox{.}}{2017}]%
        {papernot2017practical}
\bibfield{author}{\bibinfo{person}{Nicolas Papernot}, \bibinfo{person}{Patrick
  McDaniel}, \bibinfo{person}{Ian Goodfellow}, \bibinfo{person}{Somesh Jha},
  \bibinfo{person}{Z~Berkay Celik}, {and} \bibinfo{person}{Ananthram Swami}.}
  \bibinfo{year}{2017}\natexlab{}.
\newblock \showarticletitle{Practical black-box attacks against machine
  learning}. In \bibinfo{booktitle}{\emph{Proceedings of the 2017 ACM on Asia
  conference on computer and communications security}}.
  \bibinfo{pages}{506--519}.
\newblock


\bibitem[\protect\citeauthoryear{{Papernot}, {McDaniel}, {Wu}, {Jha}, and
  {Swami}}{{Papernot} et~al\mbox{.}}{2016}]%
        {distillation_SP}
\bibfield{author}{\bibinfo{person}{N. {Papernot}}, \bibinfo{person}{P.
  {McDaniel}}, \bibinfo{person}{X. {Wu}}, \bibinfo{person}{S. {Jha}}, {and}
  \bibinfo{person}{A. {Swami}}.} \bibinfo{year}{2016}\natexlab{}.
\newblock \showarticletitle{Distillation as a Defense to Adversarial
  Perturbations Against Deep Neural Networks}. In
  \bibinfo{booktitle}{\emph{2016 IEEE Symposium on Security and Privacy (SP)}}.
  \bibinfo{pages}{582--597}.
\newblock
\showISSN{2375-1207}
\urldef\tempurl%
\url{https://doi.org/10.1109/SP.2016.41}
\showDOI{\tempurl}


\bibitem[\protect\citeauthoryear{Papernot, McDaniel, Jha, Fredrikson, Celik,
  and Swami}{Papernot et~al\mbox{.}}{2015}]%
        {SMA}
\bibfield{author}{\bibinfo{person}{Nicolas Papernot},
  \bibinfo{person}{Patrick~D. McDaniel}, \bibinfo{person}{Somesh Jha},
  \bibinfo{person}{Matt Fredrikson}, \bibinfo{person}{Z.~Berkay Celik}, {and}
  \bibinfo{person}{Ananthram Swami}.} \bibinfo{year}{2015}\natexlab{}.
\newblock \showarticletitle{The Limitations of Deep Learning in Adversarial
  Settings}.
\newblock \bibinfo{journal}{\emph{CoRR}}  \bibinfo{volume}{abs/1511.07528}
  (\bibinfo{year}{2015}).
\newblock
\showeprint[arxiv]{1511.07528}
\urldef\tempurl%
\url{http://arxiv.org/abs/1511.07528}
\showURL{%
\tempurl}


\bibitem[\protect\citeauthoryear{Paszke, Gross, Massa, Lerer, Bradbury, Chanan,
  Killeen, Lin, Gimelshein, Antiga, Desmaison, Köpf, Yang, DeVito, Raison,
  Tejani, Chilamkurthy, Steiner, Fang, Bai, and Chintala}{Paszke
  et~al\mbox{.}}{2019}]%
        {PyTorch}
\bibfield{author}{\bibinfo{person}{Adam Paszke}, \bibinfo{person}{Sam Gross},
  \bibinfo{person}{Francisco Massa}, \bibinfo{person}{Adam Lerer},
  \bibinfo{person}{James Bradbury}, \bibinfo{person}{Gregory Chanan},
  \bibinfo{person}{Trevor Killeen}, \bibinfo{person}{Zeming Lin},
  \bibinfo{person}{Natalia Gimelshein}, \bibinfo{person}{Luca Antiga},
  \bibinfo{person}{Alban Desmaison}, \bibinfo{person}{Andreas Köpf},
  \bibinfo{person}{Edward Yang}, \bibinfo{person}{Zach DeVito},
  \bibinfo{person}{Martin Raison}, \bibinfo{person}{Alykhan Tejani},
  \bibinfo{person}{Sasank Chilamkurthy}, \bibinfo{person}{Benoit Steiner},
  \bibinfo{person}{Lu Fang}, \bibinfo{person}{Junjie Bai}, {and}
  \bibinfo{person}{Soumith Chintala}.} \bibinfo{year}{2019}\natexlab{}.
\newblock \bibinfo{title}{PyTorch: An Imperative Style, High-Performance Deep
  Learning Library}.
\newblock
\newblock
\showeprint[arxiv]{1912.01703}~[cs.LG]


\bibitem[\protect\citeauthoryear{Pierson and Gashler}{Pierson and
  Gashler}{2017}]%
        {pierson2017deep}
\bibfield{author}{\bibinfo{person}{Harry~A Pierson} {and}
  \bibinfo{person}{Michael~S Gashler}.} \bibinfo{year}{2017}\natexlab{}.
\newblock \showarticletitle{Deep learning in robotics: a review of recent
  research}.
\newblock \bibinfo{journal}{\emph{Advanced Robotics}} \bibinfo{volume}{31},
  \bibinfo{number}{16} (\bibinfo{year}{2017}), \bibinfo{pages}{821--835}.
\newblock


\bibitem[\protect\citeauthoryear{Raghunathan, Steinhardt, and
  Liang}{Raghunathan et~al\mbox{.}}{2018}]%
        {defense_certified}
\bibfield{author}{\bibinfo{person}{Aditi Raghunathan}, \bibinfo{person}{Jacob
  Steinhardt}, {and} \bibinfo{person}{Percy Liang}.}
  \bibinfo{year}{2018}\natexlab{}.
\newblock \bibinfo{title}{Certified Defenses against Adversarial Examples}.
\newblock
\newblock
\showeprint[arxiv]{1801.09344}~[cs.LG]


\bibitem[\protect\citeauthoryear{Redmon and Farhadi}{Redmon and
  Farhadi}{2016}]%
        {redmon2016yolo9000}
\bibfield{author}{\bibinfo{person}{Joseph Redmon} {and} \bibinfo{person}{Ali
  Farhadi}.} \bibinfo{year}{2016}\natexlab{}.
\newblock \bibinfo{title}{YOLO9000: Better, Faster, Stronger}.
\newblock
\newblock
\showeprint[arxiv]{1612.08242}~[cs.CV]


\bibitem[\protect\citeauthoryear{Ross and Doshi-Velez}{Ross and
  Doshi-Velez}{2018}]%
        {ross2018improving}
\bibfield{author}{\bibinfo{person}{Andrew~Slavin Ross} {and}
  \bibinfo{person}{Finale Doshi-Velez}.} \bibinfo{year}{2018}\natexlab{}.
\newblock \showarticletitle{Improving the adversarial robustness and
  interpretability of deep neural networks by regularizing their input
  gradients}. In \bibinfo{booktitle}{\emph{Thirty-second AAAI conference on
  artificial intelligence}}.
\newblock


\bibitem[\protect\citeauthoryear{Samangouei, Kabkab, and Chellappa}{Samangouei
  et~al\mbox{.}}{2018}]%
        {samangouei2018defense}
\bibfield{author}{\bibinfo{person}{Pouya Samangouei}, \bibinfo{person}{Maya
  Kabkab}, {and} \bibinfo{person}{Rama Chellappa}.}
  \bibinfo{year}{2018}\natexlab{}.
\newblock \showarticletitle{Defense-gan: Protecting classifiers against
  adversarial attacks using generative models}.
\newblock \bibinfo{journal}{\emph{arXiv preprint arXiv:1805.06605}}
  (\bibinfo{year}{2018}).
\newblock


\bibitem[\protect\citeauthoryear{Samek}{Samek}{2019}]%
        {samek2019explainable}
\bibfield{author}{\bibinfo{person}{Wojciech Samek}.}
  \bibinfo{year}{2019}\natexlab{}.
\newblock \bibinfo{booktitle}{\emph{Explainable AI: interpreting, explaining
  and visualizing deep learning}}. Vol.~\bibinfo{volume}{11700}.
\newblock \bibinfo{publisher}{Springer Nature}.
\newblock


\bibitem[\protect\citeauthoryear{Simonyan and Zisserman}{Simonyan and
  Zisserman}{2014}]%
        {simonyan2014deep}
\bibfield{author}{\bibinfo{person}{Karen Simonyan} {and}
  \bibinfo{person}{Andrew Zisserman}.} \bibinfo{year}{2014}\natexlab{}.
\newblock \bibinfo{title}{Very Deep Convolutional Networks for Large-Scale
  Image Recognition}.
\newblock
\newblock
\showeprint[arxiv]{1409.1556}~[cs.CV]


\bibitem[\protect\citeauthoryear{Sinha, Namkoong, and Duchi}{Sinha
  et~al\mbox{.}}{2017}]%
        {certifying}
\bibfield{author}{\bibinfo{person}{Aman Sinha}, \bibinfo{person}{Hongseok
  Namkoong}, {and} \bibinfo{person}{John Duchi}.}
  \bibinfo{year}{2017}\natexlab{}.
\newblock \bibinfo{title}{Certifying Some Distributional Robustness with
  Principled Adversarial Training}.
\newblock
\newblock
\showeprint[arxiv]{1710.10571}~[stat.ML]


\bibitem[\protect\citeauthoryear{Song, Shokri, and Mittal}{Song
  et~al\mbox{.}}{2019}]%
        {rezaShokry}
\bibfield{author}{\bibinfo{person}{Liwei Song}, \bibinfo{person}{Reza Shokri},
  {and} \bibinfo{person}{Prateek Mittal}.} \bibinfo{year}{2019}\natexlab{}.
\newblock \showarticletitle{Privacy Risks of Securing Machine Learning Models
  against Adversarial Examples}.
\newblock \bibinfo{journal}{\emph{Proceedings of the 2019 ACM SIGSAC Conference
  on Computer and Communications Security}} (\bibinfo{date}{Nov}
  \bibinfo{year}{2019}).
\newblock
\showISBNx{9781450367479}
\urldef\tempurl%
\url{https://doi.org/10.1145/3319535.3354211}
\showDOI{\tempurl}


\bibitem[\protect\citeauthoryear{Szegedy, Zaremba, Sutskever, Bruna, Erhan,
  Goodfellow, and Fergus}{Szegedy et~al\mbox{.}}{2013}]%
        {vulnerable}
\bibfield{author}{\bibinfo{person}{Christian Szegedy},
  \bibinfo{person}{Wojciech Zaremba}, \bibinfo{person}{Ilya Sutskever},
  \bibinfo{person}{Joan Bruna}, \bibinfo{person}{Dumitru Erhan},
  \bibinfo{person}{Ian Goodfellow}, {and} \bibinfo{person}{Rob Fergus}.}
  \bibinfo{year}{2013}\natexlab{}.
\newblock \bibinfo{title}{Intriguing properties of neural networks}.
\newblock
\newblock
\showeprint[arxiv]{1312.6199}~[cs.CV]


\bibitem[\protect\citeauthoryear{Tartaglione, Leps\o~y, Fiandrotti, and
  Francini}{Tartaglione et~al\mbox{.}}{2018}]%
        {NIPS2018_sparce}
\bibfield{author}{\bibinfo{person}{Enzo Tartaglione}, \bibinfo{person}{Skjalg
  Leps\o~y}, \bibinfo{person}{Attilio Fiandrotti}, {and}
  \bibinfo{person}{Gianluca Francini}.} \bibinfo{year}{2018}\natexlab{}.
\newblock \showarticletitle{Learning sparse neural networks via
  sensitivity-driven regularization}.
\newblock In \bibinfo{booktitle}{\emph{Advances in Neural Information
  Processing Systems 31}}, \bibfield{editor}{\bibinfo{person}{S.~Bengio},
  \bibinfo{person}{H.~Wallach}, \bibinfo{person}{H.~Larochelle},
  \bibinfo{person}{K.~Grauman}, \bibinfo{person}{N.~Cesa-Bianchi}, {and}
  \bibinfo{person}{R.~Garnett}} (Eds.). \bibinfo{publisher}{Curran Associates,
  Inc.}, \bibinfo{pages}{3878--3888}.
\newblock
\urldef\tempurl%
\url{http://papers.nips.cc/paper/7644-learning-sparse-neural-networks-via-sensitivity-driven-regularization.pdf}
\showURL{%
\tempurl}


\bibitem[\protect\citeauthoryear{{Tong}, {Nagle}, and {Rutenbar}}{{Tong}
  et~al\mbox{.}}{2000}]%
        {trunca}
\bibfield{author}{\bibinfo{person}{J.~Y.~F. {Tong}}, \bibinfo{person}{D.
  {Nagle}}, {and} \bibinfo{person}{R.~A. {Rutenbar}}.}
  \bibinfo{year}{2000}\natexlab{}.
\newblock \showarticletitle{Reducing power by optimizing the necessary
  precision/range of floating-point arithmetic}.
\newblock \bibinfo{journal}{\emph{IEEE Transactions on Very Large Scale
  Integration (VLSI) Systems}} \bibinfo{volume}{8}, \bibinfo{number}{3}
  (\bibinfo{date}{June} \bibinfo{year}{2000}), \bibinfo{pages}{273--286}.
\newblock
\urldef\tempurl%
\url{https://doi.org/10.1109/92.845894}
\showDOI{\tempurl}


\bibitem[\protect\citeauthoryear{Tram{\`e}r, Kurakin, Papernot, Goodfellow,
  Boneh, and McDaniel}{Tram{\`e}r et~al\mbox{.}}{2017}]%
        {tramer2017ensemble}
\bibfield{author}{\bibinfo{person}{Florian Tram{\`e}r}, \bibinfo{person}{Alexey
  Kurakin}, \bibinfo{person}{Nicolas Papernot}, \bibinfo{person}{Ian
  Goodfellow}, \bibinfo{person}{Dan Boneh}, {and} \bibinfo{person}{Patrick
  McDaniel}.} \bibinfo{year}{2017}\natexlab{}.
\newblock \showarticletitle{Ensemble adversarial training: Attacks and
  defenses}.
\newblock \bibinfo{journal}{\emph{arXiv preprint arXiv:1705.07204}}
  (\bibinfo{year}{2017}).
\newblock


\bibitem[\protect\citeauthoryear{Venceslai, Marchisio, Alouani, Martina, and
  Shafique}{Venceslai et~al\mbox{.}}{2020}]%
        {neuroattack}
\bibfield{author}{\bibinfo{person}{Valerio Venceslai}, \bibinfo{person}{Alberto
  Marchisio}, \bibinfo{person}{Ihsen Alouani}, \bibinfo{person}{Maurizio
  Martina}, {and} \bibinfo{person}{Muhammad Shafique}.}
  \bibinfo{year}{2020}\natexlab{}.
\newblock \bibinfo{title}{NeuroAttack: Undermining Spiking Neural Networks
  Security through Externally Triggered Bit-Flips}.
\newblock
\newblock
\showeprint[arxiv]{2005.08041}~[cs.CR]


\bibitem[\protect\citeauthoryear{Xie, Wu, Maaten, Yuille, and He}{Xie
  et~al\mbox{.}}{2019}]%
        {xie2019feature}
\bibfield{author}{\bibinfo{person}{Cihang Xie}, \bibinfo{person}{Yuxin Wu},
  \bibinfo{person}{Laurens van~der Maaten}, \bibinfo{person}{Alan~L Yuille},
  {and} \bibinfo{person}{Kaiming He}.} \bibinfo{year}{2019}\natexlab{}.
\newblock \showarticletitle{Feature denoising for improving adversarial
  robustness}. In \bibinfo{booktitle}{\emph{Proceedings of the IEEE Conference
  on Computer Vision and Pattern Recognition}}. \bibinfo{pages}{501--509}.
\newblock


\bibitem[\protect\citeauthoryear{Yuan, He, Zhu, Bhat, and Li}{Yuan
  et~al\mbox{.}}{2017}]%
        {pbform}
\bibfield{author}{\bibinfo{person}{Xiaoyong Yuan}, \bibinfo{person}{Pan He},
  \bibinfo{person}{Qile Zhu}, \bibinfo{person}{Rajendra~Rana Bhat}, {and}
  \bibinfo{person}{Xiaolin Li}.} \bibinfo{year}{2017}\natexlab{}.
\newblock \showarticletitle{Adversarial Examples: Attacks and Defenses for Deep
  Learning}.
\newblock \bibinfo{journal}{\emph{CoRR}}  \bibinfo{volume}{abs/1712.07107}
  (\bibinfo{year}{2017}).
\newblock
\showeprint[arxiv]{1712.07107}
\urldef\tempurl%
\url{http://arxiv.org/abs/1712.07107}
\showURL{%
\tempurl}


\bibitem[\protect\citeauthoryear{Zhou, Wu, Ni, Zhou, Wen, and Zou}{Zhou
  et~al\mbox{.}}{2018}]%
        {zhou2018dorefanet}
\bibfield{author}{\bibinfo{person}{Shuchang Zhou}, \bibinfo{person}{Yuxin Wu},
  \bibinfo{person}{Zekun Ni}, \bibinfo{person}{Xinyu Zhou}, \bibinfo{person}{He
  Wen}, {and} \bibinfo{person}{Yuheng Zou}.} \bibinfo{year}{2018}\natexlab{}.
\newblock \bibinfo{title}{DoReFa-Net: Training Low Bitwidth Convolutional
  Neural Networks with Low Bitwidth Gradients}.
\newblock
\newblock
\showeprint[arxiv]{1606.06160}~[cs.NE]


\end{thebibliography}

\end{document}